\gdef\islinuxolivetti{F}
\gdef\PSfonts{T}
\magnification\magstep1

\newdimen\papwidth
\newdimen\papheight
\newskip\beforesectionskipamount  
\newskip\sectionskipamount 
\def\sectionskip{\vskip\sectionskipamount}
\def\beforesectionskip{\vskip\beforesectionskipamount}
\papwidth=16truecm
\if F\islinuxolivetti
\papheight=22truecm
\voffset=0.4truecm
\hoffset=0.4truecm
\else
\papheight=16truecm
\voffset=-1.5truecm
\hoffset=0.4truecm
\fi
\hsize=\papwidth
\vsize=\papheight
\nopagenumbers
\headline={\ifnum\pageno>1 {\hss\tenrm-\ \folio\ -\hss} \else
{\hfill}\fi}
\newdimen\texpscorrection
\texpscorrection=0.15truecm 

\def\sectionsize{\twelvepoint}
\def\sectiontype{\bf}
\def\subsectionsize{}
\def\subsectiontype{\bf}
\def\em{\sl}
\newfam\truecmsy
\newfam\truecmr
\newfam\msbfam
\newfam\scriptfam
\newfam\truecmsy
\newskip\ttglue 
\if T\islinuxolivetti
\papheight=11.5truecm
\fi
\if F\PSfonts
\font\twelverm=cmr12
\font\tenrm=cmr10
\font\eightrm=cmr8
\font\sevenrm=cmr7
\font\sixrm=cmr6
\font\fiverm=cmr5

\font\twelvebf=cmbx12
\font\tenbf=cmbx10
\font\eightbf=cmbx8
\font\sevenbf=cmbx7
\font\sixbf=cmbx6
\font\fivebf=cmbx5

\font\twelveit=cmti12
\font\tenit=cmti10
\font\eightit=cmti8
\font\sevenit=cmti7
\font\sixit=cmti6
\font\fiveit=cmti5

\font\twelvesl=cmsl12
\font\tensl=cmsl10
\font\eightsl=cmsl8
\font\sevensl=cmsl7
\font\sixsl=cmsl6
\font\fivesl=cmsl5

\font\twelvei=cmmi12
\font\teni=cmmi10
\font\eighti=cmmi8
\font\seveni=cmmi7
\font\sixi=cmmi6
\font\fivei=cmmi5

\font\twelvesy=cmsy10	at	12pt
\font\tensy=cmsy10
\font\eightsy=cmsy8
\font\sevensy=cmsy7
\font\sixsy=cmsy6
\font\fivesy=cmsy5
\font\twelvetruecmsy=cmsy10	at	12pt
\font\tentruecmsy=cmsy10
\font\eighttruecmsy=cmsy8
\font\seventruecmsy=cmsy7
\font\sixtruecmsy=cmsy6
\font\fivetruecmsy=cmsy5

\font\twelvetruecmr=cmr12
\font\tentruecmr=cmr10
\font\eighttruecmr=cmr8
\font\seventruecmr=cmr7
\font\sixtruecmr=cmr6
\font\fivetruecmr=cmr5

\font\twelvebf=cmbx12
\font\tenbf=cmbx10
\font\eightbf=cmbx8
\font\sevenbf=cmbx7
\font\sixbf=cmbx6
\font\fivebf=cmbx5

\font\twelvett=cmtt12
\font\tentt=cmtt10
\font\eighttt=cmtt8

\font\twelveex=cmex10	at	12pt
\font\tenex=cmex10

\font\twelvemsb=msbm10	at	12pt
\font\tenmsb=msbm10
\font\eightmsb=msbm8
\font\sevenmsb=msbm7
\font\sixmsb=msbm6
\font\fivemsb=msbm5

\font\twelvescr=eusm10 at 12pt
\font\tenscr=eusm10
\font\eightscr=eusm8
\font\sevenscr=eusm7
\font\sixscr=eusm6
\font\fivescr=eusm5
\fi
\if T\PSfonts
\font\twelverm=ptmr	at	12pt
\font\tenrm=ptmr	at	10pt
\font\eightrm=ptmr	at	8pt
\font\sevenrm=ptmr	at	7pt
\font\sixrm=ptmr	at	6pt
\font\fiverm=ptmr	at	5pt

\font\twelvebf=ptmb	at	12pt
\font\tenbf=ptmb	at	10pt
\font\eightbf=ptmb	at	8pt
\font\sevenbf=ptmb	at	7pt
\font\sixbf=ptmb	at	6pt
\font\fivebf=ptmb	at	5pt

\font\twelveit=ptmri	at	12pt
\font\tenit=ptmri	at	10pt
\font\eightit=ptmri	at	8pt
\font\sevenit=ptmri	at	7pt
\font\sixit=ptmri	at	6pt
\font\fiveit=ptmri	at	5pt

\font\twelvesl=ptmro	at	12pt
\font\tensl=ptmro	at	10pt
\font\eightsl=ptmro	at	8pt
\font\sevensl=ptmro	at	7pt
\font\sixsl=ptmro	at	6pt
\font\fivesl=ptmro	at	5pt

\font\twelvei=cmmi12
\font\teni=cmmi10
\font\eighti=cmmi8
\font\seveni=cmmi7
\font\sixi=cmmi6
\font\fivei=cmmi5

\font\twelvesy=cmsy10	at	12pt
\font\tensy=cmsy10
\font\eightsy=cmsy8
\font\sevensy=cmsy7
\font\sixsy=cmsy6
\font\fivesy=cmsy5
\font\twelvetruecmsy=cmsy10	at	12pt
\font\tentruecmsy=cmsy10
\font\eighttruecmsy=cmsy8
\font\seventruecmsy=cmsy7
\font\sixtruecmsy=cmsy6
\font\fivetruecmsy=cmsy5

\font\twelvetruecmr=cmr12
\font\tentruecmr=cmr10
\font\eighttruecmr=cmr8
\font\seventruecmr=cmr7
\font\sixtruecmr=cmr6
\font\fivetruecmr=cmr5

\font\twelvebf=cmbx12
\font\tenbf=cmbx10
\font\eightbf=cmbx8
\font\sevenbf=cmbx7
\font\sixbf=cmbx6
\font\fivebf=cmbx5

\font\twelvett=cmtt12
\font\tentt=cmtt10
\font\eighttt=cmtt8

\font\twelveex=cmex10	at	12pt
\font\tenex=cmex10

\font\twelvemsb=msbm10	at	12pt
\font\tenmsb=msbm10
\font\eightmsb=msbm8
\font\sevenmsb=msbm7
\font\sixmsb=msbm6
\font\fivemsb=msbm5

\font\twelvescr=eusm10 at 12pt
\font\tenscr=eusm10
\font\eightscr=eusm8
\font\sevenscr=eusm7
\font\sixscr=eusm6
\font\fivescr=eusm5
\fi
\def\eightpoint{\def\rm{\fam0\eightrm}%
\textfont0=\eightrm
  \scriptfont0=\sixrm
  \scriptscriptfont0=\fiverm 
\textfont1=\eighti
  \scriptfont1=\sixi
  \scriptscriptfont1=\fivei 
\textfont2=\eightsy
  \scriptfont2=\sixsy
  \scriptscriptfont2=\fivesy 
\textfont3=\tenex
  \scriptfont3=\tenex
  \scriptscriptfont3=\tenex 
\textfont\itfam=\eightit
  \scriptfont\itfam=\sixit
  \scriptscriptfont\itfam=\fiveit 
  \def\it{\fam\itfam\eightit}%
\textfont\slfam=\eightsl
  \scriptfont\slfam=\sixsl
  \scriptscriptfont\slfam=\fivesl 
  \def\sl{\fam\slfam\eightsl}%
\textfont\ttfam=\eighttt
  \def\tt{\fam\ttfam\eighttt}%
\textfont\bffam=\eightbf
  \scriptfont\bffam=\sixbf
  \scriptscriptfont\bffam=\fivebf
  \def\bf{\fam\bffam\eightbf}%
\textfont\scriptfam=\eightscr
  \scriptfont\scriptfam=\sixscr
  \scriptscriptfont\scriptfam=\fivescr
  \def\script{\fam\scriptfam\eightscr}%
\textfont\msbfam=\eightmsb
  \scriptfont\msbfam=\sixmsb
  \scriptscriptfont\msbfam=\fivemsb
  \def\bb{\fam\msbfam\eightmsb}%
\textfont\truecmr=\eighttruecmr
  \scriptfont\truecmr=\sixtruecmr
  \scriptscriptfont\truecmr=\fivetruecmr
  \def\truerm{\fam\truecmr\eighttruecmr}%
\textfont\truecmsy=\eighttruecmsy
  \scriptfont\truecmsy=\sixtruecmsy
  \scriptscriptfont\truecmsy=\fivetruecmsy
\tt \ttglue=.5em plus.25em minus.15em 
\normalbaselineskip=9pt
\setbox\strutbox=\hbox{\vrule height7pt depth2pt width0pt}%
\normalbaselines
\rm
}

\def\tenpoint{\def\rm{\fam0\tenrm}%
\textfont0=\tenrm
  \scriptfont0=\sevenrm
  \scriptscriptfont0=\fiverm 
\textfont1=\teni
  \scriptfont1=\seveni
  \scriptscriptfont1=\fivei 
\textfont2=\tensy
  \scriptfont2=\sevensy
  \scriptscriptfont2=\fivesy 
\textfont3=\tenex
  \scriptfont3=\tenex
  \scriptscriptfont3=\tenex 
\textfont\itfam=\tenit
  \scriptfont\itfam=\sevenit
  \scriptscriptfont\itfam=\fiveit 
  \def\it{\fam\itfam\tenit}%
\textfont\slfam=\tensl
  \scriptfont\slfam=\sevensl
  \scriptscriptfont\slfam=\fivesl 
  \def\sl{\fam\slfam\tensl}%
\textfont\ttfam=\tentt
  \def\tt{\fam\ttfam\tentt}%
\textfont\bffam=\tenbf
  \scriptfont\bffam=\sevenbf
  \scriptscriptfont\bffam=\fivebf
  \def\bf{\fam\bffam\tenbf}%
\textfont\scriptfam=\tenscr
  \scriptfont\scriptfam=\sevenscr
  \scriptscriptfont\scriptfam=\fivescr
  \def\script{\fam\scriptfam\tenscr}%
\textfont\msbfam=\tenmsb
  \scriptfont\msbfam=\sevenmsb
  \scriptscriptfont\msbfam=\fivemsb
  \def\bb{\fam\msbfam\tenmsb}%
\textfont\truecmr=\tentruecmr
  \scriptfont\truecmr=\seventruecmr
  \scriptscriptfont\truecmr=\fivetruecmr
  \def\truerm{\fam\truecmr\tentruecmr}%
\textfont\truecmsy=\tentruecmsy
  \scriptfont\truecmsy=\seventruecmsy
  \scriptscriptfont\truecmsy=\fivetruecmsy
\tt \ttglue=.5em plus.25em minus.15em 
\normalbaselineskip=12pt
\setbox\strutbox=\hbox{\vrule height8.5pt depth3.5pt width0pt}%
\normalbaselines
\rm
}

\def\twelvepoint{\def\rm{\fam0\twelverm}%
\textfont0=\twelverm
  \scriptfont0=\tenrm
  \scriptscriptfont0=\eightrm 
\textfont1=\twelvei
  \scriptfont1=\teni
  \scriptscriptfont1=\eighti 
\textfont2=\twelvesy
  \scriptfont2=\tensy
  \scriptscriptfont2=\eightsy 
\textfont3=\twelveex
  \scriptfont3=\twelveex
  \scriptscriptfont3=\twelveex 
\textfont\itfam=\twelveit
  \scriptfont\itfam=\tenit
  \scriptscriptfont\itfam=\eightit 
  \def\it{\fam\itfam\twelveit}%
\textfont\slfam=\twelvesl
  \scriptfont\slfam=\tensl
  \scriptscriptfont\slfam=\eightsl 
  \def\sl{\fam\slfam\twelvesl}%
\textfont\ttfam=\twelvett
  \def\tt{\fam\ttfam\twelvett}%
\textfont\bffam=\twelvebf
  \scriptfont\bffam=\tenbf
  \scriptscriptfont\bffam=\eightbf
  \def\bf{\fam\bffam\twelvebf}%
\textfont\scriptfam=\twelvescr
  \scriptfont\scriptfam=\tenscr
  \scriptscriptfont\scriptfam=\eightscr
  \def\script{\fam\scriptfam\twelvescr}%
\textfont\msbfam=\twelvemsb
  \scriptfont\msbfam=\tenmsb
  \scriptscriptfont\msbfam=\eightmsb
  \def\bb{\fam\msbfam\twelvemsb}%
\textfont\truecmr=\twelvetruecmr
  \scriptfont\truecmr=\tentruecmr
  \scriptscriptfont\truecmr=\eighttruecmr
  \def\truerm{\fam\truecmr\twelvetruecmr}%
\textfont\truecmsy=\twelvetruecmsy
  \scriptfont\truecmsy=\tentruecmsy
  \scriptscriptfont\truecmsy=\eighttruecmsy
\tt \ttglue=.5em plus.25em minus.15em 
\setbox\strutbox=\hbox{\vrule height7pt depth2pt width0pt}%
\normalbaselineskip=15pt
\normalbaselines
\rm
}
%
\fontdimen16\tensy=2.7pt
\fontdimen13\tensy=4.3pt
\fontdimen17\tensy=2.7pt
\fontdimen14\tensy=4.3pt
\fontdimen18\tensy=4.3pt
\fontdimen16\eightsy=2.7pt
\fontdimen13\eightsy=4.3pt
\fontdimen17\eightsy=2.7pt
\fontdimen14\eightsy=4.3pt
\fontdimen18\eightsy=4.3pt
%
\def\hexnumber#1{\ifcase#1 0\or1\or2\or3\or4\or5\or6\or7\or8\or9\or
 A\or B\or C\or D\or E\or F\fi}
\mathcode`\=="3\hexnumber\truecmr3D
\mathchardef\not="3\hexnumber\truecmsy36
\mathcode`\+="2\hexnumber\truecmr2B
\mathcode`\(="4\hexnumber\truecmr28
\mathcode`\)="5\hexnumber\truecmr29
\mathcode`\!="5\hexnumber\truecmr21
\mathcode`\(="4\hexnumber\truecmr28
\mathcode`\)="5\hexnumber\truecmr29

\def\bar{\mathaccent"0\hexnumber\truecmr16 }

\def\Phi{\mathchar"0\hexnumber\truecmr08 }
\def\Gamma {\mathchar"0\hexnumber\truecmr00 }
\def\Delta {\mathchar"0\hexnumber\truecmr01 }
\def\Theta {\mathchar"0\hexnumber\truecmr02 }
\def\Lambda{\mathchar"0\hexnumber\truecmr03 }
\def\Xi {\mathchar"0\hexnumber\truecmr04 }
\def\Pi{\mathchar"0\hexnumber\truecmr05 }
\def\Sigma{\mathchar"0\hexnumber\truecmr06 }
\def\Upsilon {\mathchar"0\hexnumber\truecmr07 }
\def\Phi {\mathchar"0\hexnumber\truecmr08 }
\def\Psi {\mathchar"0\hexnumber\truecmr09 }
\def\Omega{\mathchar"0\hexnumber\truecmr0A }
\newcount\EQNcount \EQNcount=1
\newcount\CLAIMcount \CLAIMcount=1
\newcount\SECTIONcount \SECTIONcount=0
\newcount\SUBSECTIONcount \SUBSECTIONcount=1
\def\ifff(#1,#2,#3){\ifundefined{#1#2}%
\expandafter\xdef\csname #1#2\endcsname{#3}\else%
\immediate\write16{!!!!!doubly defined #1,#2}\fi}
\def\NEWDEF #1,#2,#3 {\ifff({#1},{#2},{#3})}
\def\actualnumber{\number\SECTIONcount}
\def\EQ(#1){\lmargin(#1)\eqno\tag(#1)}
\def\NR(#1){&\lmargin(#1)\tag(#1)\cr}  
\def\tag(#1){\lmargin(#1)({\rm \actualnumber}.\number\EQNcount)
 \NEWDEF e,#1,(\actualnumber.\number\EQNcount)
\global\advance\EQNcount by 1
}
\def\SECT(#1)#2\par{\lmargin(#1)\SECTION#2\par
\NEWDEF s,#1,{\actualnumber}
}
\def\SUBSECT(#1)#2\par{\lmargin(#1)
\SUBSECTION#2\par 
\NEWDEF s,#1,{\actualnumber.\number\SUBSECTIONcount}
}
\def\CLAIM #1(#2) #3\par{
\vskip.1in\medbreak\noindent
{\lmargin(#2)\bf #1\ \actualnumber.\number\CLAIMcount.} {\sl #3}\par
\NEWDEF c,#2,{#1\ \actualnumber.\number\CLAIMcount}
\global\advance\CLAIMcount by 1
\ifdim\lastskip<\medskipamount
\removelastskip\penalty55\medskip\fi}
\def\CLAIMNONR #1(#2) #3\par{
\vskip.1in\medbreak\noindent
{\lmargin(#2)\bf #1.} {\sl #3}\par
\NEWDEF c,#2,{#1}
\global\advance\CLAIMcount by 1
\ifdim\lastskip<\medskipamount
\removelastskip\penalty55\medskip\fi}
\def\SECTION#1\par{\vskip0pt plus.3\vsize\penalty-75
    \vskip0pt plus -.3\vsize
    \global\advance\SECTIONcount by 1
    \beforesectionskip\noindent
{\sectionsize\sectiontype \actualnumber.\ #1}
    \EQNcount=1
    \CLAIMcount=1
    \SUBSECTIONcount=1
    \nobreak\sectionskip\noindent}
\def\SECTIONNONR#1\par{\vskip0pt plus.3\vsize\penalty-75
    \vskip0pt plus -.3\vsize
    \global\advance\SECTIONcount by 1
    \beforesectionskip\noindent
{\sectionsize\sectiontype  #1}
     \EQNcount=1
     \CLAIMcount=1
     \SUBSECTIONcount=1
     \nobreak\sectionskip\noindent}
\def\SUBSECTION#1\par{\vskip0pt plus.2\vsize\penalty-75%
    \vskip0pt plus -.2\vsize%
    \beforesectionskip\noindent%
{\subsectionsize\subsectiontype \actualnumber.\number\SUBSECTIONcount.\ #1}
    \global\advance\SUBSECTIONcount by 1
    \nobreak\sectionskip\noindent}
\def\SUBSECTIONNONR#1\par{\vskip0pt plus.2\vsize\penalty-75
    \vskip0pt plus -.2\vsize
\beforesectionskip\noindent
{\subsectionsize\subsectiontype #1}
    \nobreak\sectionskip\noindent\noindent}
\def\ifundefined#1{\expandafter\ifx\csname#1\endcsname\relax}
\def\equ(#1){\ifundefined{e#1}$\spadesuit$#1\else\csname e#1\endcsname\fi}
\def\clm(#1){\ifundefined{c#1}$\spadesuit$#1\else\csname c#1\endcsname\fi}
\def\sec(#1){\ifundefined{s#1}$\spadesuit$#1
\else Section \csname s#1\endcsname\fi}
\let\endarg=\par
\def\finish{\def\endarg{\par\endgroup}}
\def\start{\endarg\begingroup}

 \def\beginFROM{\start\parskip=0pt\vskip\baselineskip
\def\finish{\def\endarg{\egroup\par\endgroup}}
  \vbox\bgroup\obeylines\eightpoint\em\finish}

\def\ABSTRACT#1\par{
\vskip 1in {\noindent\sectionsize\sectiontype Abstract.} #1 \par}

\def\TODAY{\number\day~\ifcase\month\or January \or February \or March \or
April \or May \or June
\or July \or August \or September \or October \or November \or December \fi
\number\year\timecount=\number\time
\divide\timecount by 60
}
\newcount\timecount
\def\DRAFT{\def\lmargin(##1){\strut\vadjust{\kern-\strutdepth
\vtop to \strutdepth{
\baselineskip\strutdepth\vss\rlap{\kern-1.2 truecm\eightpoint{##1}}}}}
\font\footfont=cmti7
\footline={{\footfont \hfil File:\jobname, \TODAY,  \number\timecount h}}
}
\newbox\strutboxJPE
\setbox\strutboxJPE=\hbox{\strut}
\def\subitem#1#2\par{\vskip\baselineskip\vskip-\ht\strutboxJPE{\item{#1}#2}}
\gdef\strutdepth{\dp\strutbox}
\def\lmargin(#1){}
\def\period{\unskip.\spacefactor3000 { }}
%
%
\newbox\noboxJPE
\newbox\byboxJPE
\newbox\paperboxJPE
\newbox\yrboxJPE
\newbox\jourboxJPE
\newbox\pagesboxJPE
\newbox\volboxJPE
\newbox\preprintboxJPE
\newbox\toappearboxJPE
\newbox\bookboxJPE
\newbox\bybookboxJPE
\newbox\publisherboxJPE
\newbox\inprintboxJPE
\def\refclearJPE{
   \setbox\noboxJPE=\null             \gdef\isnoJPE{F}
   \setbox\byboxJPE=\null             \gdef\isbyJPE{F}
   \setbox\paperboxJPE=\null          \gdef\ispaperJPE{F}
   \setbox\yrboxJPE=\null             \gdef\isyrJPE{F}
   \setbox\jourboxJPE=\null           \gdef\isjourJPE{F}
   \setbox\pagesboxJPE=\null          \gdef\ispagesJPE{F}
   \setbox\volboxJPE=\null            \gdef\isvolJPE{F}
   \setbox\preprintboxJPE=\null       \gdef\ispreprintJPE{F}
   \setbox\toappearboxJPE=\null       \gdef\istoappearJPE{F}
   \setbox\inprintboxJPE=\null        \gdef\isinprintJPE{F}
   \setbox\bookboxJPE=\null           \gdef\isbookJPE{F}  \gdef\isinbookJPE{F}
     
   \setbox\bybookboxJPE=\null         \gdef\isbybookJPE{F}
   \setbox\publisherboxJPE=\null      \gdef\ispublisherJPE{F}
     
}
\def\widestlabel#1{\setbox0=\hbox{#1\enspace}\refindent=\wd0\relax}
\def\ref{\refclearJPE\bgroup}
\def\no   {\egroup\gdef\isnoJPE{T}\setbox\noboxJPE=\hbox\bgroup}
\def\by   {\egroup\gdef\isbyJPE{T}\setbox\byboxJPE=\hbox\bgroup}
\def\paper{\egroup\gdef\ispaperJPE{T}\setbox\paperboxJPE=\hbox\bgroup}
\def\yr{\egroup\gdef\isyrJPE{T}\setbox\yrboxJPE=\hbox\bgroup}
\def\jour{\egroup\gdef\isjourJPE{T}\setbox\jourboxJPE=\hbox\bgroup}
\def\pages{\egroup\gdef\ispagesJPE{T}\setbox\pagesboxJPE=\hbox\bgroup}
\def\vol{\egroup\gdef\isvolJPE{T}\setbox\volboxJPE=\hbox\bgroup\bf}
\def\preprint{\egroup\gdef
\ispreprintJPE{T}\setbox\preprintboxJPE=\hbox\bgroup}
\def\toappear{\egroup\gdef
\istoappearJPE{T}\setbox\toappearboxJPE=\hbox\bgroup}
\def\inprint{\egroup\gdef
\isinprintJPE{T}\setbox\inprintboxJPE=\hbox\bgroup}
\def\book{\egroup\gdef\isbookJPE{T}\setbox\bookboxJPE=\hbox\bgroup\em}
\def\publisher{\egroup\gdef
\ispublisherJPE{T}\setbox\publisherboxJPE=\hbox\bgroup}
\def\inbook{\egroup\gdef\isinbookJPE{T}\setbox\bookboxJPE=\hbox\bgroup\em}
\def\bybook{\egroup\gdef\isbybookJPE{T}\setbox\bybookboxJPE=\hbox\bgroup}
\newdimen\refindent
\refindent=5em
\def\endref{\egroup \sfcode`.=1000
 \if T\isnoJPE
 \hangindent\refindent\hangafter=1
      \noindent\hbox to\refindent{[\unhbox\noboxJPE\unskip]\hss}\ignorespaces
     \else  \noindent    \fi
 \if T\isbyJPE    \unhbox\byboxJPE\unskip: \fi
 \if T\ispaperJPE \unhbox\paperboxJPE\unskip\period \fi
 \if T\isbookJPE {\it\unhbox\bookboxJPE\unskip}\if T\ispublisherJPE, \else.
\fi\fi
 \if T\isinbookJPE In {\it\unhbox\bookboxJPE\unskip}\if T\isbybookJPE,
\else\period \fi\fi
 \if T\isbybookJPE  (\unhbox\bybookboxJPE\unskip)\period \fi
 \if T\ispublisherJPE \unhbox\publisherboxJPE\unskip \if T\isjourJPE, \else\if
T\isyrJPE \  \else\period \fi\fi\fi
 \if T\istoappearJPE (To appear)\period \fi
 \if T\ispreprintJPE Pre\-print\period \fi
 \if T\isjourJPE    \unhbox\jourboxJPE\unskip\ \fi
 \if T\isvolJPE     \unhbox\volboxJPE\unskip\if T\ispagesJPE, \else\ \fi\fi
 \if T\ispagesJPE   \unhbox\pagesboxJPE\unskip\  \fi
 \if T\isyrJPE      (\unhbox\yrboxJPE\unskip)\period \fi
 \if T\isinprintJPE (in print)\period \fi
\filbreak
}
\def\hexnumber#1{\ifcase#1 0\or1\or2\or3\or4\or5\or6\or7\or8\or9\or
 A\or B\or C\or D\or E\or F\fi}
\textfont\msbfam=\tenmsb
\scriptfont\msbfam=\sevenmsb
\scriptscriptfont\msbfam=\fivemsb
\mathchardef\varkappa="0\hexnumber\msbfam7B
\newcount\FIGUREcount \FIGUREcount=0
\newdimen\figcenter
\def\fig(#1){\ifundefined{fig#1}%
\global\advance\FIGUREcount by 1%
\NEWDEF fig,#1,{Fig.\ \number\FIGUREcount}
\immediate\write16{ FIG \number\FIGUREcount : #1}
\fi
\csname fig#1\endcsname\relax}
\def\figure #1 #2 #3 #4\cr{\null%
\ifundefined{fig#1}%
\global\advance\FIGUREcount by 1%
\NEWDEF fig,#1,{Fig.\ \number\FIGUREcount}
\immediate\write16{  FIG \number\FIGUREcount : #1}
\fi
{\goodbreak\figcenter=\hsize\relax
\advance\figcenter by -#3truecm
\divide\figcenter by 2
\midinsert\vskip #2truecm\noindent\hskip\figcenter
\includegraphics{#1}\vskip 0.8truecm\noindent \vbox{\eightpoint\noindent
{\bf\fig(#1)}: #4}\endinsert}}
\def\figurewithtex #1 #2 #3 #4 #5\cr{\null%
\ifundefined{fig#1}%
\global\advance\FIGUREcount by 1%
\NEWDEF fig,#1,{Fig.\ \number\FIGUREcount}
\immediate\write16{ FIG \number\FIGUREcount: #1}
\fi
{\goodbreak\figcenter=\hsize\relax
\advance\figcenter by -#4truecm
\divide\figcenter by 2
\midinsert\vskip #3truecm\noindent\hskip\figcenter
\includegraphics{#1}{\hskip\texpscorrection\input #2 }\vskip 0.8truecm\noindent \vbox{\eightpoint\noindent
{\bf\fig(#1)}: #5}\endinsert}}
\def\figurewithtexplus #1 #2 #3 #4 #5 #6\cr{\null%
\ifundefined{fig#1}%
\global\advance\FIGUREcount by 1%
\NEWDEF fig,#1,{Fig.\ \number\FIGUREcount}
\immediate\write16{ FIG \number\FIGUREcount: #1}
\fi
{\goodbreak\figcenter=\hsize\relax
\advance\figcenter by -#4truecm
\divide\figcenter by 2
\midinsert\vskip #3truecm\noindent\hskip\figcenter
\includegraphics{#1}{\hskip\texpscorrection\input #2 }\vskip #5truecm\noindent \vbox{\eightpoint\noindent
{\bf\fig(#1)}: #6}\endinsert}}
\catcode`@=11
\def\footnote#1{\let\@sf\empty 
  \ifhmode\edef\@sf{\spacefactor\the\spacefactor}\/\fi
  #1\@sf\vfootnote{#1}}
\def\vfootnote#1{\insert\footins\bgroup\eightpoint
  \interlinepenalty\interfootnotelinepenalty
  \splittopskip\ht\strutbox 
  \splitmaxdepth\dp\strutbox \floatingpenalty\@MM
  \leftskip\z@skip \rightskip\z@skip \spaceskip\z@skip \xspaceskip\z@skip
  \textindent{#1}\footstrut\futurelet\next\fo@t}
\def\fo@t{\ifcat\bgroup\noexpand\next \let\next\f@@t
  \else\let\next\f@t\fi \next}
\def\f@@t{\bgroup\aftergroup\@foot\let\next}
\def\f@t#1{#1\@foot}
\def\@foot{\strut\egroup}
\def\footstrut{\vbox to\splittopskip{}}
\skip\footins=\bigskipamount 
\count\footins=1000 
\dimen\footins=8in 
\catcode`@=12 

\def\BB{{\script B}}
\def\CC{{\script C}}
\def\EE{{\script E}}

\def\OO{{\script O}}

\def\VV{{\script V}}
\def\HALF{{\textstyle{1\over 2}}}

\def\QED{\hfill\smallskip
         \line{$\hfill{\vcenter{\vbox{\hrule height 0.2pt
	\hbox{\vrule width 0.2pt height 1.8ex \kern 1.8ex
		\vrule width 0.2pt}
	\hrule height 0.2pt}}}$
               \ \ \ \ \ \ }
         \bigskip}
\def\real{{\bf R}}

\def\integer{{\bf Z}}

\def\Im{{\rm Im\,}}
\def\PROOF{\medskip\noindent{\bf Proof.\ }}
\def\REMARK{\medskip\noindent{\bf Remark.\ }}
\def\LIKEREMARK#1{\medskip\noindent{\bf #1.\ }}
\tenpoint
\normalbaselineskip=5.25mm
\baselineskip=5.25mm
\parskip=10pt
\beforesectionskipamount=24pt plus8pt minus8pt
\sectionskipamount=3pt plus1pt minus1pt
\def\em{\it}
\normalbaselineskip=12pt
\baselineskip=12pt
\parskip=0pt
\parindent=22.222pt
\beforesectionskipamount=24pt plus0pt minus6pt
\sectionskipamount=7pt plus3pt minus0pt
\overfullrule=0pt
\hfuzz=2pt
\nopagenumbers
\headline={\ifnum\pageno>1 {\hss\tenrm-\ \folio\ -\hss} \else
{\hfill}\fi}
\if F\PSfonts
\font\titlefont=cmbx10 scaled\magstep2

\font\toplinefont=cmr10
\font\pagenumberfont=cmr10
\let\tenpoint=\rm
\else
\font\titlefont=ptmb at 14 pt

\font\toplinefont=cmcsc10
\font\pagenumberfont=ptmb at 10pt
\fi
\newdimen\itemindent\itemindent=1.5em

\def\textindent#1{\indent\llap{#1\enspace}\ignorespaces}
\def\item{\par\noindent
\hangindent\itemindent\hangafter=1\relax
\setitemmark}
\def\setitemindent#1{\setbox0=\hbox{\ignorespaces#1\unskip\enspace}%
\itemindent=\wd0\relax
\message{|\string\setitemindent: Mark width modified to hold
         |`\string#1' plus an \string\enspace\space gap. }%
}
\def\setitemmark#1{\checkitemmark{#1}%
\hbox to\itemindent{\hss#1\enspace}\ignorespaces}
\def\checkitemmark#1{\setbox0=\hbox{\enspace#1}%
\ifdim\wd0>\itemindent
   \message{|\string\item: Your mark `\string#1' is too wide. }%
\fi}
\def\SECTION#1\par{\vskip0pt plus.2\vsize\penalty-75
    \vskip0pt plus -.2\vsize
    \global\advance\SECTIONcount by 1
    \beforesectionskip\noindent
{\sectionsize\sectiontype \actualnumber.\ #1}
    \EQNcount=1
    \CLAIMcount=1
    \SUBSECTIONcount=1
    \nobreak\sectionskip\noindent}

\headline
{\ifnum\pageno>1 {\toplinefont Extensive Properties of the Ginzburg-Landau Equation}
\hfill{\pagenumberfont\folio}\fi}
\let\epsilon=\varepsilon
\let\phi=\varphi
\def\per{{\rm per}}
\def\GG{{\cal G}}
\def\UU{{\script U}}
\def\FF{{\script F}}
\def\SS{{\script S}}

\def\ee/{$\epsilon $-entropy}
\def\const{{\rm const.}}
\def\Linfty{{\rm L}^\infty}
\def\D{{\rm D}}
\def\B{{\bf B}}
\let\para=\S
\def\S{{\bf S}}
\def\BS{{\bf B}\setminus{\bf S}}
\def\logtwo{\log}
\def\kernint{\!\!}
\def\CC{{\cal C}}
\let\truett=\tt
\fontdimen3\tentt=2pt\fontdimen4\tentt=2pt
\def\tt{\hfill\break\null\kern -2truecm\truett **** }
\setitemindent{iii)}
\rm
\vskip 3cm
\centerline{\titlefont Extensive Properties of the}
\vskip 0.5cm
\centerline{\titlefont Complex Ginzburg-Landau Equation}
\vskip 3cm
\centerline{\twelvepoint Pierre Collet\footnote{$^{1}$}
{Centre de Physique Th\'eorique, Laboratoire CNRS UPR 14,
Ecole Polytechnique, F-91128 Palaiseau Cedex (France).}
and Jean-Pierre Eckmann\footnote{$^{2}$}
{Dept.~de Physique Th\'eorique et Section de
Math\'ematiques, Universit\'e de Gen\`eve,
CH-1211 Gen\`eve 4 (Suisse).}}
\vskip 3cm
{\eightpoint \baselineskip 10pt
\noindent{\bf Abstract}: We study the set of solutions
of the complex Ginzburg-Landau equation
in $\real^d$, $d<3$. We consider the global attracting set ({\it i.e.}, the
forward map of the set of bounded initial data), and restrict it
to a cube $Q_L$ of side $L$. We cover this set by a (minimal)
number $N_{Q_L}(\epsilon )$ of balls of radius $\epsilon $ in
$\Linfty(Q_L)$.
We show that the Kolmogorov \ee/ per unit length,
$H_\epsilon =\lim_{L\to\infty} L^{-d} \logtwo N_{Q_L}(\epsilon
)$ exists. In particular, we bound $H_\epsilon $ by
$\OO\bigl(\logtwo(1/\epsilon )\bigr)$, which shows that the attracting set is {\em
smaller} than the set of bounded analytic functions in a strip.
We finally give a positive lower bound:
$H_\epsilon>\OO\bigl (\logtwo(1/\epsilon )\bigr ) $.}

\SECTION Introduction

In the last few years, considerable effort has been made towards a
better understanding of partial differential equations of parabolic
type in {\em infinite space}. A typical equation is for example the
complex Ginzburg-Landau equation (CGL) on  $\real ^d$:
$$
\partial_t A\,=\,(1+i\alpha)\Delta A+A-(1+i\beta)A|A|^2~.
\EQ(CGL)
$$
Such equations show, at least numerically, in certain parameter
ranges, interesting 
``chaotic'' behavior, and our aim here is to discuss notions of {\em
chaoticity per unit length} for such systems. Our discussion will be
restricted to the CGL, but it will become clear from the methods of
the proofs that the results can be extended without much additional
work to other problems in which high frequencies are strongly damped.

A first idea which comes to mind in the context of measuring
chaoticity
is the notion of~
``dimension per unit length.'' As we shall see, this quantity is
a well-defined and useful concept in dynamical
systems with finite-dimensional phase space. While the ``standard''
definition leads to infinite dimensions for finite segments of
infinite systems, we shall see that an adequate definition, first
introduced by Kolmogorov and Tikhomirov [KT], leads to finite bounds
which measure the ``complexity'' of the set under study.

\SECTION Attracting Sets

In the study of PDE's, there are several definitions of
``attractors.'' In this work, we concentrate our attention onto
attracting sets (which may be larger than attractors).

\LIKEREMARK{Definition}A set $\GG$ is called an {\em attracting
set} with fundamental neighborhood $\UU$ for the flow $\Phi_t$ if
\item{i)}The set $\GG$ is compact.
\item{ii)}For every open set $\VV\supset \GG$ we have $\Phi_t
\UU\subset \VV$ when $t$ is large enough.
\item{iii)}The set $\GG$ is invariant.

The open set $\cup_{t>0}(\Phi_t)^{-1}(\UU)$ is called the basin of
attraction of $\GG$. If the basin of attraction is the full space,
then $\GG$ is called a {\em global attracting set}.

\REMARK One finds a large number of definitions of ``attractors'' in
the literature [T], [MS]. Our terminology is inspired from the theory of
dynamical systems. In particular, an attracting set is {\em not} an
attractor in the sense of dynamical systems,
it is usually larger than the attractor. See
also [ER] for a discussion of these issues.

We will consider
the
Eq.\equ(CGL) in a (large) box $Q_L$ of side $L$ in $\real^d$, with
periodic boundary
conditions. Let $\GG_{Q_L}$ denote the global attracting set for this problem.
It has been shown [GH] that $\GG_{Q_L}$ is a compact set in 
$\Linfty
_{\per,Q_L}$
(since the set
is made up of functions analytic in a strip around the real axis).

For the CGL on the {\em infinite space} the situation is somewhat more
complicated. A non-trivial invariant set
$\GG$ can be defined in the topology of uniformly continuous
functions as follows: First, if $B$ is a large enough ball of
uniformly continuous functions in
$\Linfty$,
there is a
finite time $T_0(B)$ such that for any $T>T_0(B)$ one has
$$
\Theta^{T}(B)\,\subset\, B~,
$$
where $t\mapsto\Theta^t$ is the flow defined by the CGL.
The set $\GG(B,T)$ is then defined by
$$
{\GG(B,T)}\,=\,\bigcap_{n\ge0}\Theta^{nT}(B)~.
\EQ(gdef)
$$
It can be shown (see [MS]) that this set is invariant and that it does
not depend on the initial ball $B$ (if it is large enough) nor on the
(large enough) time $T>T_0(B)$. Thus, we {\em define} $\GG= \GG(B,T)$.
It is
made up of functions which extend to bounded analytic functions in a
strip. Its width and the bound on the functions
only depend on the parameters of the problem.
These facts can be found scattered in the literature, but are
``well-known,''
see, {\it e.g.}, [C].
The set $\GG$ probably lacks
properties i) and ii) above in the topology of uniformly continuous functions.
We will nevertheless call it a globally
attracting set since in [MS] it was proven that in local and/or weaker
topologies conditions of the type of i) and ii) are satisfied.
The set $\GG$ defined by Eq.\equ(gdef) will be our main object of study.

\SECTION Dimension in Finite Volume

We define $M_{Q_L}(\epsilon )$ to be the minimum number of balls of
radius $\epsilon $ in
$\Linfty_{\per,Q_L} $ needed to cover $\GG_{Q_L}$. One can then define
$$
\CC_{Q_L}\,=\, \limsup _{\epsilon \to 0} {\logtwo M_{Q_L}(\epsilon )\over \logtwo
(1/\epsilon )}~.
$$
The technical term [M, 5.3] for this is the ``upper Minkowski
dimension.'' This dimension is an upper bound for the Hausdorff
dimension. It is also equal to the (upper) box-counting dimension (in
which the positions of the boxes are centered on a dyadic grid).

It has been shown by Ghidaglia and H\'eron [GH]
that $\CC_{Q_L}$ satisfies an ``extensive bound:''
\CLAIM Proposition(ext1) For CGL one has in dimensions $d=1,2$, the bound
$$
\limsup_{L\to\infty} {\CC_{Q_L}\over L^d} \,<\,\infty ~.
\EQ(GH)
$$

To our knowledge, it is an open problem to show the existence of the
limit in \equ(GH). The difficulty in obtaining a proof is that the
familiar methods of statistical mechanics of matching together pieces
of configurations to obtain a subadditivity bound of the form (written
for simplicity for the case of dimension $d=1$ and with $L$ instead of
$Q_L$)
$$
\CC_{{L_1+L_2}} \,\le\, \CC_{L_1}+\CC_{L_2}+\OO(1)~,
$$
do not seem to work.

One can try to define a sort of ``local'' dimension by restricting the
global problem to a local window. But this idea does not work either
as we show now:
For example
consider the global attracting set $\GG$ for CGL
on the
{\em infinite line}. As we have said before, this set is compact in
a local topology which is not too fine.
Take again a cube $Q_L$ of side $L$ in $\real^d$
and then denote by $N_{Q_L}(\epsilon )$ the minimum number of balls of
radius $\epsilon $ in
$\Linfty (Q_L)$ needed
to cover $\GG|_{Q_L}$. Again, this number is finite. But we have the
following
\CLAIM Lemma(counter) For every $L>0$ we have
$$
\liminf_{\epsilon \to 0} {\logtwo N_{Q_L}(\epsilon )\over \logtwo (1/\epsilon)}
\,=\,
\infty ~.
\EQ(diverge1)
$$

\REMARK In other words, this lemma shows that the lower Minkowski
dimension for the restriction of $\GG$ to $Q_L$ is infinite. Thus, there
are many more functions in $\GG|_{Q_L}$ than in $\GG_{Q_L}$.
In fact, our proof will show a little more, namely
\CLAIM Corollary(haus) The Hausdorff dimension of $\GG|_{Q_L}$ is
infinite for every $L>0$.

\PROOF The proof will be given in Section 5.

The example of \clm(counter) and \clm(haus)
teaches us that the restriction to nice functions on the infinite line
produces ``too many''
functions on a {\em finite} interval, as the observation (the
$\epsilon $) becomes infinitely accurate.
This fact calls for a new
kind of definition. Such a possibility is offered by the
considerations of Kolmogorov and Tikhomirov [KT].

\SECTION The \ee/ per Unit Length

The basic idea is to take the limit of infinite $L$ {\em
before} considering the behavior as $\epsilon $ goes to zero. Thus,
with the definitions of the preceding section, we now define
$$
H_\epsilon \,=\, \lim_{L\to\infty } {\logtwo N_{Q_L}(\epsilon )\over L^d}~.
$$
In the paper [KT], this quantity was studied for different sets of
functions.
The authors considered in particular three classes of functions on the
real line: 

\item{i)}The class $\EE_\sigma(C)$ of entire functions $f$ which are
bounded
by $|f(z)|\le C e^{\sigma |\Im z|}$.
\item{ii)}The class $\FF_{p,\sigma}(C)$ of entire functions $f$ with growth
of order $p>1$, which are bounded by
$|f(z)|\le C e^{\sigma |\Im z|^p}$.
\item{iii)}The class $\SS_h(C)$ of bounded analytic functions in the
strip $|\Im z|< h$ with a bound $|f(z)|<C$.

\medskip\noindent
For these classes the following result holds
\CLAIM Theorem(TK) [KT]. One has the bounds:
$$
\eqalign{
H_\epsilon \,\sim\,\cases
{
(2\sigma/\pi)\cdot\logtwo(1/\epsilon )& for the class $\EE_\sigma(C)$,\cr
{2\sigma ^{1/p} p^2\over
\pi (2p-1) (p-1)^{1-1/p}}\cdot \bigl (\logtwo(1/\epsilon )\bigr )^{2-1/p}& for the class $\FF_{p,\sigma}(C)$,\cr
{1\over \pi h} \cdot\bigl (\logtwo(1/\epsilon )\bigr )^2
& for the class $\SS_h(C)$,\cr
}
}
$$
as $\epsilon \to 0$ in the sense that the l.h.s~divided by the
r.h.s~has limit equal to 1.

\LIKEREMARK{Notation}It will sometimes be convenient to write the
dependence on the space such as $H_\epsilon \bigl (\EE_\sigma(C)\bigr )$.

Our main result is the following
\CLAIM Theorem(everything) The global attracting set $\GG$ of CGL
satisfies a bound
$$
H_\epsilon(\GG )\,\le\, \const \logtwo(1/\epsilon )~,
\EQ(aim)
$$
where the constant depends only on the parameters of the equation.

\REMARK The reader should note
that this result contains new information
about the set $\GG$ of limiting states.
It is for example well known that the solutions of CGL are
analytic and bounded in a strip, that is, they are in the class $\SS_h(C)$ for some
$h>0$ and some $C<\infty $.
This alone, however would only give a bound
$$
{1\over \pi h\log2 e} \bigl (\logtwo(1/\epsilon )\bigr )^2~,
$$
as we have seen in \clm(TK).
Therefore, \clm(everything) shows that the long-time
solutions are not only analytic in a strip,
but form a proper subset of $\SS_h(C)$
with smaller \ee/ per unit length. On the other hand, the set $\GG$ is
in general {\em not contained in} the class $\EE_\sigma(C)$, because some
stationary solutions are not entire. For example for the real
Ginzburg-Landau equation, the function $\tanh(x/\sqrt{2})$ is a
stationary solution with a singularity in the complex plane.
For the CGL, Hocking and Stewartson [HS, Eq.(5.2)] describe
time-periodic
solutions which exist in certain parameter ranges of $\alpha $ and
$\beta $, and which are again not entire in $x$ and are of the form
$$
\const e^{i a_1 t} {\rm sech} (a_2 x)^{1+i a_3}~,
$$
where $a_i=a_i(\alpha ,\beta )$ can be found in [HS].

\SECTION Proof of \clm(counter) and \clm(haus)

We fix $L>0$, and we want to show that
$$
\liminf_{\epsilon\to0}{\logtwo N_{Q_L}(\epsilon)\over \logtwo(1/\epsilon)}
\,=\,\infty~,
\EQ(counter)
$$
where $N_{Q_L}(\epsilon )$ is the minimum number of balls needed to cover
$\GG|_{Q_L}$.

The idea of the proof is to observe that $\GG|_{Q_L}$ contains subsets
of arbitrarily high Hausdorff dimension. These subsets are essentially
parts of the unstable manifold of the 0 solution.

We begin, as in [GH],
by considering periodic solutions of period $\Lambda $ for
various $\Lambda $. In that space, for $\Lambda $ large enough, the origin is
an
unstable fixed point and the spectrum of the generator for
the linearized evolution is
$$
\biggl \{ 1 - (1+i\alpha ){4 \pi^2\over  \Lambda ^2}
(n_1^2+\cdots+ n_d^2)~\biggl |~ n_i\in \integer \biggr \}~.
$$
Thus, the origin is a hyperbolic fixed point
if $2\pi/\Lambda $ is irrational. In
that case the local unstable manifold $W$ of the origin has dimension
$D_\Lambda \equiv\OO(1)\Lambda ^d$.
In other words, we have a $\CC^1$ map $\Psi_\Lambda $
from a neighborhood $U$ of $0$
in $\real ^{D_\Lambda }$ to $W$ which is injective (and in fact has
differentiable inverse). This construction can be justified in a
Sobolev space with sufficiently high index [GH, Remark 3.2, p. 289], [G].

This unstable manifold is of course contained in the global
attracting set $\GG_\Lambda $. But it is also in $\GG$.
We can consider $W$ as a subset $\Phi(U)$ in $\GG $ and look
at it in $\Linfty(Q_L)$ (with $L\ll \Lambda $). We would like to prove
that there also it has 
a dimension equal to $D_\Lambda $. Note that there is a $\CC^1$ map
$\Phi$ which maps $U$ to $W$.
We claim $\Phi$ is injective. Indeed, assume
not, then there are two different
points $u_1$ and $u_2$ in $U$ such that on $Q_L$ the functions
$\Phi(u_1)$ and $\Phi(u_2)$ coincide.
But since these functions are analytic in a strip
they coincide everywhere and hence $u_1=u_2$:
we have a contradiction.

This implies that in $\Linfty(Q_L)$, the local unstable manifold
$W$ has also dimension
$D_\Lambda$. Therefore, for $\epsilon$ small enough, we need at least
$$
\left({1\over \epsilon}\right)^{D_\Lambda -1}
$$
balls of radius $\epsilon$ to cover it. The assertion \equ(counter)
follows by letting
$\Lambda $ tend to infinity. The proof of \clm(counter) is complete.
Since we have constructed a lower bound for every $L$, the \clm(haus)
follows at once.

\SECTION Upper Bound on the \ee/ per Unit Length

We study in this section the quantity $H_\epsilon (\GG )$ for the global
attracting set on $\GG$ for the CGL on the whole space.
We begin by
\CLAIM Theorem(1) For fixed $\epsilon>0$, the sequence $\logtwo
N_{Q_L}(\epsilon)/L^d$ has a limit when $L$ goes to infinity, and there exists
a constant $C$ such that
$$
\lim_{L\to\infty}{\logtwo N_{Q_L}(\epsilon)\over L^d}\,\le\,
C\logtwo(1/\epsilon)
{}~.
\EQ(1)
$$
The constant $C$ only depends on the parameters of the CGL.

We first prove the existence of the limit.

\CLAIM Lemma(2) For any fixed $\epsilon>0$, the sequence $\logtwo
N_{Q_L}(\epsilon)/L^d$ has a limit when $L$ goes to infinity.

\PROOF Let $B$ and $B'$ denote two disjoint bounded sets of $\real^d$.
We denote by $N_B(\epsilon)$ the minimum number of balls
in $\Linfty(B)$ of radius $\epsilon$ which is needed to cover $\GG|_B $.
Since we are using the sup norm, it is easy to verify that
$$
N_{B\cup B'}(\epsilon)\,\le\, N_B(\epsilon)N_{B'}(\epsilon)~,
\EQ(subadd)
$$
because one can choose the functions in $B$ and $B'$ independently.
The lemma follows by the standard sub-additivity argument, see [R],
since the $Q_L$ form a van Hove sequence.

We now begin working towards a bound
relating $N_{Q_L}(\epsilon)$ and $N_{Q_L}(\epsilon/2)$. The bound
will be inefficient for small $L$ but becomes asymptotically better.
We let the CGL semi-flow act on balls in $\Linfty(Q_L)$,
and we will analyze the
deformation of these balls by looking at the difference between the
trajectory of the center and the trajectory of the other points.

We begin by considering functions $f$ and $g$, both in $\GG$.
We set
$$
w_0\,=\,g-f~.
$$
It is
left to the reader to verify that there are bounded functions $R$ and
$S$ of space and time such that
$$
\partial_t w\,=\,(1+i\alpha)\Delta w+Rw+S\overline w~,
\EQ(CGLD)
$$
more precisely, we set $w(t=0)= w_0$, and
$$
R\,=\,1-(1+i\beta )(f+g)\bar f~,\quad
S\,=\,-(1+i\beta )g^2~.
$$
Note that since $\GG$ is bounded in a suitable space of analytic functions,
there is a constant $K>1$ which
depends only on $\alpha$ and $\beta$ such that
$$
\sup_t\|w(t,\,\cdot\,)\|_\infty+\sup_t\|\nabla
w(t,\,\cdot\,)\|_\infty\,\le\, K~.
\EQ(wbound)
$$
We want to show that if $w_0$ is small in $Q_L$ then the same is true
for
the solution of \equ(CGLD) up to time 1.
This might seem not to be true because a large perturbation may reach
$Q_L$ from
the outside. However, using localization techniques, we now show that
this effect can only take place near the boundary.

We will therefore introduce a
layer of width $\ell$ near the boundary of the cube $Q_L$, and we
assume $\ell<L$.
We assume  $Q_L$ to be centered at the origin and
consider the cube $Q_{L-\ell}$ also centered at the origin.

We use as in [CE] the family of space cut-off
functions
$$
\varphi_a(x)\,=\,Z{1\over (1+|x-a|^4)^d}\,\equiv\, \phi(x-a)     ~,
$$
where
$$
Z^{-1}\,=\,\int dx\,{1\over (1+|x|^4)^d}~.
$$

\CLAIM Lemma(3) Let $f$ and $g$ be in $\GG$, and let $w_0=f-g$.
In dimension $d\le3$, if $\ell>1/\epsilon$ and $w$ is a
solution of Eq.\equ(CGLD) with initial data $w_0$ satisfying
$$
\|w_0\|_\infty\,\le\, 2K~,\qquad\hbox{\rm and}\qquad
\sup_{x\in Q_L}|w_0(x)|\,\le\,\epsilon~,
$$
then
$$
\sup_{0\le t\le 1}\sup_{a\in Q_{L-\ell }}
\int dx\,\varphi_a(x)\,
 |w(t,x)|^2\,\le\,\OO(\epsilon^2)~,
\EQ(93)
$$
$$
\sup_{0\le t\le 1}\sup_{a\in Q_{L-\ell }}
 |w(t,a)|\,\le\,\OO (\epsilon )~,
\EQ(99)
$$
and
$$
\sup_{x\in Q_{L-\ell }}|\nabla w(t=1 ,x)|\,\le\,\OO(\epsilon )~.
\EQ(100)
$$
These bounds depend on $K$ but are independent of $\ell>1/\epsilon $.

\REMARK The constant $K=K(\alpha ,\beta )$ in this lemma is the one found in
Eq.\equ(wbound). Below, the notation $\OO_{\alpha ,\beta }(1)$ will stand for a
bound
which depends only on $\alpha $, $\beta $ and this $K(\alpha ,\beta
)$, but not on $L$, $\ell$ or $\epsilon $.

\PROOF We begin by bounding $X\equiv\partial_t\int
dx\,\varphi_a(x)|w(t,x)|^2$. Using Eq.\equ(CGLD) we have:
$$
X\,=\,\int dx\, \bar w \phi_a \bigl ((1+i\alpha)\Delta
w+Rw+S\overline w\bigr ) + cc~,
$$
where $cc$ denotes the complex conjugate. Integrating by parts we get
$$
\eqalign{
X\,&=\,-(1+i\alpha )\int dx\,\phi_a |\nabla w|^2
-(1+i\alpha )\int dx\, \bar w (\nabla \phi_a\cdot \nabla w)
\cr
\,&+\, \int dx\,\phi_a \bar w \bigl (Rw +S \bar w\bigr )+ cc~.\cr
}
\EQ(91)
$$
By the choice of $\phi_a$ we have $|\nabla \phi_a(x)|\le \const \phi_a(x)$,
uniformly in $x$ and $a$. Therefore $X$ can be bounded above by
$$
X\,\le\, -2\int dx\, \phi_a |\nabla w|^2
+\OO_{\alpha ,\beta }(1)\int dx\,\phi_a |w| |\nabla w|
+\OO_{\alpha ,\beta }(1)\int dx\, \phi_a |w|^2~.
$$
By polarization, and using that $\phi_a>0$, we get a bound
$$
\partial_t\int dx\,\phi_a |w|^2\,\le\, -\int dx\, \phi_a |\nabla
w|^2+\OO_{\alpha ,\beta }(1) \int dx\, \phi_a |w|^2
{}~.
\EQ(95)
$$
Therefore we see that there is a
constant $C$ which depends only on
$\alpha$ and $\beta$, for which we have the differential inequality
$$
\partial_t\int dx\,\varphi_a(x)|w(t,x)|^2\,\le\, C
\int dx\,\varphi_a(x)\,|w(t,x)|^2~.
\EQ(92)
$$
Since $w(0,x)$ is bounded on $\real^d$
and small on $Q_L$, we have for
$\ell>1/\epsilon$,
$$
\sup_{a\in Q_{L-\ell }}\int dx\,\phi_a(x)|w(0,x)|^2\,\le\,
\OO(1+K^2)\epsilon^2~.
$$
To see this, split the integration region into $Q_L$ and
$\real^d\setminus Q_L$. In the first region, $w$ is small and in the
second region the integral of $\phi_a$ is small and $|w|\le K$.
Using Eq.\equ(92), we find
$$
\sup_{0\le t\le 1}\sup_{a\in Q_{L-\ell }}\int
dx\,\varphi_a(x)\,
|w(t,x)|^2\,\le\,e^{C }\OO_{\alpha ,\beta }(1)\epsilon^2\,=\,\OO_{\alpha ,\beta
}(1)\epsilon ^2~.
$$
Thus we have shown Eq.\equ(93).

We next bound the solutions in $\Linfty $.
Let $G_t$ denote the convolution kernel of the semigroup generated
by the operator $(1+i\alpha)\Delta$.
We have
$$
w(t,\,\cdot\,)\,=\, G_t\star w_0+\int_0^t\kernint ds\,
G_{t-s}\star\bigl (R(s,\,\cdot\,) w(s,\,\cdot\,)+S(s,\,\cdot\,)\bar
 w(s,\,\cdot\,)\bigr )~.
\EQ(96)
$$
We first bound the term
$$
Y_{t,s}\,\equiv\,
G_{t-s}\star \bigl (R(s,\,\cdot\,) w(s,\,\cdot\,)\bigr )~.
$$
We rewrite it as
$$
Y_{t,s}(x)\,=\,\int dy\,{G_{t-s} (x-y) \over \sqrt{\phi(x-y)}}\,
\sqrt{\phi(x-y)} \,R(s,y) w(s,y)~.
$$
By the Schwarz inequality, we get a bound
$$
Y_{t,s}^2\,\le\,\int dy\,{|G_{t-s}|^2 (x-y) \over{\phi(x-y)}}\,\cdot
\OO_{\alpha ,\beta }(1) \int dz\,\phi_x(z) |w(s,z)|^2~.
\EQ(97)
$$
Using Eq.\equ(93), the second factor in \equ(97) is bounded by
$\OO(\epsilon ^2)$.
The complex heat kernel $G$ can be bounded as follows:
\CLAIM Lemma(heat) For every $n>0$ there is a constant $C_n$ such that
$$
|G_t(z)|\,\le\,  {C_n\over (1+z^2/t)^{n/2}} \,{1\over t^{d/2}}~,
\EQ(hor)
$$
and
$$
|\nabla G_{t}(z)|\,\le\, {1\over t^{d/2}}{C_{n}\over (1+z^{2}/t)^{n}}
{|z|\over t}~.
\EQ(hor2)
$$

\PROOF Use the stationary phase method [H].

Using this lemma, the first factor in \equ(97) is bounded for $t-s<1$
and for
$n$ large enough, by
$$
\int dy\,{|G_{t-s}|^2 (x-y) \over{\phi(x-y)}}
\,\le\,
\int dy {C_n\over \bigl (1+{(x-y)^2\over t-s}\bigr )^{n}} \,{1\over
(t-s)^{d}}(1+|x-y|^4)^d\,\le\,\OO\bigl ((t-s)^{-d/2}\bigr ) ~.
$$
Inserting in \equ(97),
and integrating over $s$, we get the bound
$$
\int_0^t\kernint ds\,Y_{t,s}\,\le\, \OO(\epsilon )~,
$$
provided $d<4$. The term involving $S$ is bounded in the same
manner. The inhomogeneous term in \equ(96) is bounded by splitting the
convolution integral into the regions $y\in Q_L$ and $y\in \real^d
\setminus Q_L$. The first term gives a small contribution because
$w_0$ is $\OO(\epsilon )$ on $Q_L$ and the second contribution is
small because the kernel $G_t$ is small for $x\in Q_{L-\ell}$ and $y
\in \real^d
\setminus Q_L$. This proves Eq.\equ(99).

It remains to show Eq.\equ(100). We have
$$
\nabla w(t,\,\cdot\,)\,=\,\nabla G_t\star w_0+\int_0^tds\,
\nabla G_{t-s}\star \bigl (R(s,\,\cdot\,) w(s,\,\cdot\,)+S(s,\,\cdot\,)
\bar w(s,\,\cdot\,)\bigr )~.
$$
We deal first with the inhomogeneous term.
Using the same splitting as before, and \clm(heat), we get
$$
\sup_{x\in Q_{L-\ell }}|\bigl ((\nabla G_{t=1})\star w_0\bigr
)(x)|\,\le\,
\OO (\epsilon )~.
$$
The homogeneous term $I$ involving $R$ is:
$$
I\,=\,\int _0^t\kernint ds\, \nabla G_{t-s} \star \bigl (w R\bigr )~.
$$
We want to bound $I$ for $t=1$ and rewrite it as
$$
I\,=\,\int _0^{1/2} \!\!ds\, \nabla G_{1-s} \star \bigl (w R\bigr )
+\int _{1/2}^1 \!\!ds\, G_{1-s} \star \bigl (w\nabla R\bigr )
+\int _{1/2}^1 \!\!ds\, G_{1-s} \star \bigl (R\nabla w\bigr )\,=\,
I_1+I_2+I_3~.
$$
The term $I_2$ is bounded in the same way as the integral of
$Y_{t,s}$. To bound the term $I_1$
we observe that there is no singularity in the kernel \equ(hor2),
since $s<\HALF$, and
furthermore,
$$
|\nabla G_{1-s}(z)|\,\le\,\const \phi(z)~.
$$
Then the Schwarz inequality and the results on $w$ yield
$$
I_1\,\le\,\OO(\epsilon )~.
\EQ(100i1)
$$
Finally, consider $I_3$.
Integrating Eq.\equ(95) over $s$ from 0 to $\HALF$, we have
$$
\eqalign{
\int_0^{1/2} \!\!ds&\, \int dx\,\varphi_a(x)\,|\nabla w(s,x)|^2\,\cr
\,&\le\,
\OO(1)\int_0^{1/2}\!\! ds\, \int dx\,\varphi_a(x)\,| w(s,x)|^2 + \int dx\,
\varphi_a(x)\,| w(0,x)|^2~.\cr
}
$$
Our previous bounds show that the r.h.s.~is bounded by $\OO(\epsilon
^2)$. Therefore there is a value of $s^*\in(0,\HALF)$ for which
$$
 \int dx\,\varphi_a(x)\,|\nabla w(s^*,x)|^2\,\le\,\OO(\epsilon ^2)~.
\EQ(sstern)
$$
Furthermore, we have
\CLAIM Lemma(more) We have the bounds
$$
\partial_t \int dx\,\varphi_a(x)\,|\nabla w(t,x)|^2\,\le\,
\OO(1) \int dx\,\varphi_a(x)\,|\nabla w(t,x)|^2
+\OO(1)\int dx\,\varphi_a(x)\,|w(t,x)|^2~.
\EQ(103)
$$

\PROOF We start with
$$
\eqalign{
\partial_t \int dx\,\varphi_a\,|\nabla w|^2\,&=\,
\int dx\,\varphi_a\,\nabla \bar w\cdot \partial_t \nabla w +cc\cr
\,&=\,
\int dx\,\varphi_a\,\nabla \bar w\cdot  \nabla \biggl((1+i\alpha
)\Delta w +R w +S \bar w\biggr) +cc\cr
\,&=\,-\int dx\,\varphi_a\,\Delta \bar w  \biggl((1+i\alpha
)\Delta w +R w +S \bar w\biggr) \cr
&~~-\int dx\,(\nabla\varphi_a\,\cdot\nabla \bar w) \biggl((1+i\alpha
)\Delta w +R w +S \bar w\biggr) +cc~.\cr
}
$$
Using again the explicit form of $\phi_a$, completing the square and
polarization, as in the proof of Eq.\equ(92), the assertion follows.

We continue with the proof of \clm(3).
Let $s\in(\HALF,1]$ and $T_s=\int dx \,\phi_a(x) |\nabla w(s,x)|^2$.
Then we integrate the differential inequality
\equ(103) which reads $\partial_t T_t\le \OO(1) T_t + \OO(\epsilon
^2)$
from $s^*$ to $s$. This yields, using \equ(sstern),
$$
\eqalign
{
\int dx \,\phi_a(x) |\nabla w(s,x)|^2\,&\le\,
\exp \bigl (\OO(1) (s-s^*)\bigr )\int dx \,\phi_a(x) |\nabla
w(s^*,x)|^2+\OO(\epsilon ^2)
\cr
\,&\le\,\OO(\epsilon ^2)~.\cr
}
\EQ(104)
$$
Using this bound, we rewrite
$$
I_3\,=\,\int_{1/2}^1 \!\!ds\, \int dy\,
{G_{1-s}(x-y)\over \sqrt{\phi(x-y)}} \,{\sqrt{\phi(x-y)}}
R(s,y) \nabla w (s,y)~.
$$
Using the Schwarz inequality as in Eq.\equ(97), we get a bound
$$
I_3\,\le\, \OO(\epsilon )~.
$$
Combining the bounds on $I_1$, $I_2$ and $I_3$ completes the proof of
Eq.\equ(100). The proof of \clm(3) is complete.

\clm(3) gives us control over the evolution of differences in $\GG$,
when they are small in $\GG|_{Q_L}$. We shall now use this information
to study the deformation of balls covering $\GG|_{Q_L}$.
To formulate the next result we need the following notation:
Consider the universal attracting set $\GG$.
The quantity $N_B^{(t)}(\epsilon )$ denotes the number of balls of
radius $\epsilon $ needed to cover the set $\Theta^t(\GG)|_B$, in
$\Linfty(B)$,
where
$\Theta^t $ is the semi-flow defined by the CGL equation.
\CLAIM Proposition(balls) There are constants $c<\infty $ and
$D$, $D_1<\infty $ such that for all
sufficiently small
$\epsilon>0 $ and all $L>3/\epsilon $ one has the bound
$$
N_{Q_L}^{(t+1)}(\epsilon/2 )\,\le\,
\left (
{c\over \epsilon }\right )^{D_1 L^{d-1} \epsilon ^{-(1+d)}} D^{L^d}
N_{Q_L}^{(t)}(\epsilon )~.
\EQ(thebound)
$$

Before we prove this proposition, we need a geometric lemma:
\CLAIM Lemma(cover) Let $Q$ be a set of diameter $r$ in $\real^d$
and assume that $\FF$ is a family of complex functions $f$ on $Q$ which
satisfy the bounds
$$
|f|\,\le\,a~,\quad |\nabla f|\,\le\, b~,
$$
with $br\le c/2$.
Then one can cover $\FF$ with not more than
$$
(4a/c)^2
$$
balls of radius $c$ in $\Linfty(Q) $.

\PROOF On a disk in $\real^d$ of diameter $r$, the function $f$
varies no more than $br$ which is bounded by $c/2$. On the other hand,
one can find a set $\SS$ of $(4a/c)^2$ complex numbers of modulus less than $a$
such that every complex number of modulus less than $a$ is within
$c/2$ of $\SS$. Since $f$ varies less than $c/2$ one can find a
constant function $f^*$ with value in $\SS$ such that  $\sup_Q
|f-f^*|<c$.

\LIKEREMARK{Proof of \clm(balls)}By definition we can find, for every $t\ge0$,
$N_{Q_L}^{(t)}(\epsilon)$ balls of radius $\epsilon$ in
$\Linfty(Q_L)$ which cover $\Theta^t(\GG)|_{Q_L}$.
Therefore we can find a
collection $\BB$ of $N_{Q_L}^{(t)}(\epsilon)$ balls of radius $2\epsilon$ in
$L^\infty(Q_L)$  {\em with center in
$\Theta^t(\GG)|_{Q_L}$},
which cover $\Theta^t(\GG)|_{Q_L}$.
Let $B$ be a ball ({\it i.e.}, an element of $\BB$). We denote by
$
B\cap\Theta^t(\GG)
$ those functions in $\Theta^t(\GG)$ whose restriction to $Q_L$ is in $B$.
We have obviously $\cup_{B\in {\BB}} \bigl( B \cap \Theta^t(\GG)\bigr) \supset
\Theta^t(\GG)$, and
therefore
$$
\Theta^{t+1}(\GG)|_{Q_L}\subset\bigcup_{B\in {\BB}}\Theta^1\bigl( B \cap
\Theta^t(\GG)\bigr)|_{Q_L}~.
$$
Thus, we can move the time forward by one
unit without changing the set we cover. This will be the crux of our
argument, which will use the smoothing properties of $\Theta^1$
described in \clm(3).

We are going to cover every set $\Theta^{1}\bigl( B \cap
\Theta^t(\GG)\bigr)|_{Q_L}$
by balls of radius
$\epsilon/2$ in $\Linfty (Q_{L})$. Counting all these balls will give
the result. So we fix a $B\in{\cal B}$ and consider $\Theta^{1}\bigl( B
\cap \Theta^t(\GG)\bigr)|_{Q_L}$.
Since $B\in\BB$, its center $f$
is in $\Theta^t(\GG)|_{Q_L}$, and, since $\Theta^t(\GG)\subset\GG$,
we also have $f\in \GG|_{Q_L}$. (In fact $f$ is the restriction of a
function in $\GG$ to $Q_L$.)
Let $g$ be an arbitrary point in $\bigl (B\cap\Theta^t(\GG)\bigr )|_{Q_L}$.
Our construction makes sure that both $f$ and $g$ satisfy the
assumptions of \clm(3) (with $2\epsilon $ instead of $\epsilon $).
{}From \clm(3), there are constants $c_{1}$ and $c_{2}$ (which do not
depend on $\epsilon$, $f$, or $g$)
such that in $Q_{L-\ell }$ the following holds:
If $w_0=g-f$ and $w=\Theta^1(g)-\Theta^1(f)$, then
$$
|w|\,\le\, c_{1}\epsilon~,\qquad |\nabla w|\,\le\, c_{2}\epsilon~.
$$
Let
$$
r_1\,=\,\min\bigl (1,1/(4c_{2})\bigr )~.
$$
We partition $Q_{L-\ell }$ into disjoint
cubes $Q$ of side $r_1$ (except at the boundary
where we take possibly a strip of smaller cubes if necessary). In each
of these cubes we can apply \clm(cover) with $c=\epsilon/2$ since
$$
c_{2}\epsilon r_1\,\le\,\epsilon/4~.
$$
Therefore we can cover the
restriction of $\Theta^{1}\bigl( B \cap \Theta^t(\GG)\bigr)$ to each cube $Q$
by
$$
(4c_{1}\epsilon/(\epsilon/2))^{2}\,=\,64 c_{1}^{2}
$$
balls of radius $\epsilon/2$ in $\Linfty (Q)$.
We shall now use the same method in the corridor $Q_{L}\backslash Q_{L-\ell }$
but with balls at a different scale.
In $Q_L\backslash Q_{L-\ell}$ we have only inequality \equ(wbound) and not a
bound
$\OO(\epsilon )$ as in $Q_{L-\ell}$. Therefore
we define
$$
r_2\,=\,\epsilon/(4K)~,
$$
and again $c=\epsilon/2$. This leads to
$$
Kr_2\,=\,c/2~.
$$
We now cover the corridor $Q_{L}\backslash Q_{L-\ell }$ by cubes $Q'$
of side $r_2$
(again a smaller strip at the boundary may be needed). In each of these
cubes $Q'$ the \clm(cover) applies and we can cover $\Theta^{1}\bigl( B \cap
\Theta^t(\GG)\bigr)$
restricted to these cubes by
$$
64 K^{2}\epsilon^{-2}
$$
balls of radius $\epsilon/2$ in $\Linfty (Q')$.
We now have a covering of $Q_{L}$ by disjoint cubes. If we have a ball
of radius $\epsilon/2$ in $\Linfty $ in each cube, this defines a ball
in $\Linfty (Q_{L})$ since in $\Linfty $ the product of two
independent covers is a cover of the union of the sets, see Eq.\equ(subadd).

To get a covering of  $\Theta^{1}\bigl (B\cap\Theta^t(\GG)\bigr )$ in $\Linfty
(Q_{L})$
we have to consider all these possible balls and in particular count
them. It is easy to verify that the number of such balls is bounded by
$$
(64 c_{1}^{2})^{(1+(L-\ell )/\min(1,1/4c_{2}))^{d}}
(64 K^{2}\epsilon^{-2})^{2(1+4KL/\epsilon)^{d-1}(1+4K\ell/\epsilon)}~,
\EQ(terrible)
$$
and the inequality \equ(thebound) follows.
The proof of \clm(balls) is complete.

\LIKEREMARK{Proof of \clm(1)}Finally, we can prove \clm(1), and hence
also \clm(everything). We use \clm(balls) recursively 
by starting at time $t=1$ with $\epsilon =1$. For this case, we can
apply \clm(cover) with $a=K$, $b=K$, $r=1/(4K)$ to get
$$
N_{Q_L}^{(t=1)}(\epsilon =1)\,\le\,e^{ \OO_{\alpha ,\beta }\bigl
((2L+1)^d\bigr )}~,
$$
and using inequality \equ(thebound) inductively,
we get
$$
N_{Q_L}^{(n+1)}(2^{-n})\,\le\,
e^{\OO_{\alpha ,\beta }\bigl ((2L+1)^d\bigr )}
 D^{n L^d } \prod _{j=0}^{n-1} (2^j c)^ {D_1 L^{d-1}
2^{j(d+1)}}~.
$$
Taking logarithms and dividing by $(2L)^d$ we get
$$
{\logtwo N_{Q_L}^{(n+1)}\over (2L)^d}\,\le\,
n \logtwo D + L^{-d}\OO_{\alpha ,\beta }(L^d) + L^{-d} \OO_{\alpha ,\beta
}( n L^{d-1} 2^{n(d+1)})~.
$$
Clearly, \clm(1) follows by taking $n$ as the integer
part
of $\logtwo(1+1/\epsilon
)$.
\REMARK As asserted, $D$ only depends on the parameters of the
equation, as can be seen from Eq.\equ(terrible):
$$
D\,=\,\OO\bigl (64 c_1^{1/\min(1,1/(4c_2))}\bigr )~,
$$
where $c_1$ and $c_2$ can be found in the proof of \clm(3).
Note also that there is a crossover point (for our bound) between the behavior
described in \clm(everything), and the divergence described in
\equ(diverge1), at about $\epsilon = L^{-1/(1+d)}$.

\SECTION Lower Bound on the \ee/ per Unit Length

In this section, we construct a lower bound on $H_\epsilon(\GG )$.
The idea is to construct a subset of the ``local unstable manifold'' of
the origin with large enough \ee/ per unit length.
Working in space dimension 1 is enough, because such solutions are also
solutions (in $\Linfty$) in higher dimensions which do not depend on the other
variables (of course the lower bounds are not very accurate).
The main result of this section is then
\CLAIM Theorem(lowerbound) There is a constant $A>0$ such that for
sufficiently small $\epsilon >0$, the \ee/ per unit length of the unstable
manifold of 0 (and hence of the global attracting set $\GG$ of CGL)
satisfies the bound
$$
H_\epsilon(\GG )\,\ge\, A \logtwo(1/\epsilon )~.
\EQ(aim2)
$$

\SUBSECTION The idea of the proof

To obtain a lower bound on the \ee/ (always per unit length), we
exhibit a large enough set of
functions for which we prove that they are in the global attracting
set.
This set is built by observing that the 0 solution $u=0$ has an
unstable linear subspace which is made up of functions with momenta $k$ in
$[-1,1]$.
For these functions to be in the strongly unstable region, we restrict
our attention to the class $\EE_b(\eta)$ with $b=1/3$
of entire functions in $z=x+iy$ which are
bounded by $|f(z)|\le \eta e^{b |\Im z| }$.
The Fourier transform $\widehat f$ of a function $f$
in this class is a distribution with support in $[-b,b]$ (see [S]) and
is
therefore strongly unstable.
Furthermore, by \clm(TK) we have the bound
$$
\lim_{\epsilon\to0} {H_\epsilon \bigl (\EE_b (\eta)\bigr )\over
{2b\over \pi} \logtwo(1/\epsilon )} \,=\,1~,
\EQ(hepsilon)
$$
so there are ``many'' such functions.
(See [KT], Theorem XXII and beginning of \para 3).

We want to use the set $\EE_b(\eta)$ as the starting point for the
construction of a set in $\GG$ with positive \ee/. Thus, we want to
evolve $\EE_b(\eta)$ {\em forward} in time to reach $\GG$, using the
evolution operator $\Theta^t$ defined above. However, this would move
us far away from the solution 0 and we would lose control of the
non-linearity.
To overcome this difficulty, we first evolve the set $\EE_b(\eta)$
{\em backward} in time by a linearized evolution.
Thus, we use the method known from the usual construction of unstable
manifolds, adapted to the case of continuous spectrum.

We begin by defining the linear evolution.
Given $T>0$ we let
$\widehat\Theta_0^{T}(k)=e^{(1-(1+i\alpha)k^{2})T}$ and then
$$
\widehat{(\Theta^T_0f)}(k)\,=\,\widehat\Theta^T_0(k)\widehat f (k)\,=\,e^{(1-(1+i\alpha)k^{2})T}\widehat f(k)~.
$$
Note that the map (in $x$-space) $\Theta^T_0 : f\mapsto \Theta^T_0f$ is the evolution generated
by the linearized CGL.
Inspired by scattering theory, we will then consider the quantity
$$
S(f)\,=\,\lim_{T\to\infty}\Theta^T\Theta^{-T}_0(f)~.
$$
Since we consider the unstable manifold of 0 and stay in a vicinity of
$f=0$, the nonlinearities
should be negligible and thus the following result
seems very natural:
\CLAIM Theorem(superlower) Let $b=1/3$.
There is an $\eta_*>0$ such
that for
$\eta\le\eta_*$ the
following limit exists in $\Linfty({\real})$ for $f\in\EE_b(\eta)$:
$$
S(f)\,=\,\lim_{T\to\infty}\Theta^{T}\Theta_0^{-T}f~.
$$
Moreover,
$$
S(f)\,=\,f+Z(f)~,
$$
where $Z$ is Lipshitz continuous
in $f$, with a Lipshitz constant of order $\OO(\eta)$.

In other words, $S$ is close to the identity.
Using this kind of information, we shall see that if
two functions are separated by $\epsilon $ the functions $S(f)-S(f')$
are separated almost as much. Therefore, knowing that the set
$\EE_b(\eta)$
of $f$
has positive \ee/ implies that the set $S(\EE_b(\eta))$---which is in the
global attracting set---also has positive \ee/, as we shall show later.

\SUBSECTION The regularized linear evolution

In this subsection, we construct a somewhat more regular representation
of $\Theta_0^T$, which is needed because we consider negative $T$.

We consider the class $\EE_b(\eta)$, with $b=1/3$.
It is clear from the Paley-Wiener-Schwartz [S]
theorem that the functions $f\in\EE_b(\eta)$
have a Fourier transform $\widehat f(k)=\int
dx \,e^{ikx} f(x)$ which
is a distribution with support in $[-b,b]$.
If $\widehat f$ were a function, we could freely go back and forth
between
$k$-space and $x$-space.
To deal with this problem, we use a regularizing device.
Let $c>b$ and let $\widehat \psi$ be a positive
$\CC^\infty $ function with support in $[-c,c]$ and equal to 1 on
$[-b,b]$. We shall take $b=1/3$, $c=1/\sqrt{3}$.
Clearly $\widehat \psi (k) \widehat f(k)= \widehat f(k)$ (as a distribution) and
therefore
$\psi\star f = f$ (in $x$-space), where $\star$ denotes the
convolution product.
We define a regularized linear evolution kernel
$$
g_{T}(x)\,=\,\int dk\, e^{ikx} \widehat\psi(k) e^{T(1-(1+i\alpha
)k^2)}~,
$$
and then we define
$$
\bigl (\Theta_{0,\psi}^T f\bigr )(x)\,\equiv\,\bigl (g_{T}\star f\bigr ) (x)~.
$$
This is our regularized representation of the linear evolution.
By construction, it has the property:
If $f\in\EE_b(\eta)$, then
$$
\Theta_{0,\psi}^T f\,=\,\Theta_0^T f~,
\EQ(equal)
$$
as a distribution. But, as we shall see below, the l.h.s.\ is a well
defined function and thus we can use either of the definitions,
whichever is more convenient. Henceforth, we use the notation
$f_t$ for $\Theta^t_0f=\Theta_{0,\psi}^T f$.

\SUBSECTION Proof of the first part of \clm(superlower)

This theorem is relatively conventional, but tedious, to prove. We will
therefore only sketch the standard estimates and describe in detail
only the general sequence of estimates which are needed.

We begin the proof of the first part of \clm(superlower) with a study
of $\Theta^t$.
First we would like to prove that $\Theta^{t}(f_{-T})-f_{t-T}$
remains small for $0\le t\le T$.
\CLAIM Lemma(lower1) For $\eta$ small enough, there is a $\rho>0$
such that for any $T>0$ and any $t\in[0,T]$ we have for all
$f\in\EE_b(\eta)$,
the bound
$$
\|\Theta^{t}(f_{-T})-f_{t-T}\|_\infty \,\le\, \eta^{2} e^{-\rho(T-t)}~.
$$

\PROOF First observe that by assumption $\|f\|_\infty \le \eta$.
By definition, we have
$$
\bigl(\Theta^{-T}_0 f\bigr )(x)\,=\,\int dy\,
dk\, e^{ik(x-y)} e^{-T(1-k^2(1+i\alpha ))} \widehat \psi(k) f(y)~.
$$
Since $\widehat\psi$ is smooth and supported in $|k|\le c$, we get
from this the easy but useful bound
$$
\|f_{-T}\|_{\infty}\,\le\, \OO(\eta) e^{-(1-c^{2})T}~.
\EQ(firstlinfty)
$$
Using Eq.\equ(equal), we see that
$\Theta_{0,\psi}^{t}f_{-T}=f_{t-T}$ satisfies
$$
\partial_{t}f_{t-T}\,=\,(1+i\alpha)\partial_{x}^{2}f_{t-T}+f_{t-T}~.
$$
We let $v=\Theta^{t}(f_{-T})-f_{t-T}$, and then we find
$$
\partial_{t}v\,=\,(1+i\alpha)\partial_{x}^{2}v+v-(1+i\beta)
(v+f_{t-T})|(v+f_{t-T})|^{2}~.
$$
We write this as an integral equation using $v(0,x)=0$. We get
$$
v(t,\cdot)\,=\,-(1+i\beta)\int_{0}^{t}\kernint ds\,\Theta_{0,\psi}^{t-s}\bigl(
(v(s,\cdot)+f_{s-T})\cdot|v(s,\cdot)+f_{s-T}|^{2}\bigr)~.
\EQ(vequ)
$$
In particular there is an inhomogeneous term
$$
-(1+i\beta)\int_{0}^{t}\kernint ds\,\Theta_{0,\psi}^{t-s}\bigl(
f_{s-T}|f_{s-T}|^{2}\bigr)~.
\EQ(inhomog)
$$
This term can be bounded by using Eq.\equ(firstlinfty) and the
bound $\|\Theta^\tau_0 g\|_\infty \le e^\tau \|g\|_\infty $ (which
follows from \clm(heat)). We get
$$
\eta^{3} \OO (1)\int_{0}^{t}\kernint ds\, e^{T-s} e^{-3(1-c^{2})(T-s)}\,\le\,
\OO(\eta^{3})  ~e^{-\rho(T-t)}~,
\EQ(contra)
$$
and here the restriction on the choice of $c$ implies
$$
\rho\,=\,3(1-c^{2})-1>0~.
$$
Thus we have bounded the inhomogeneous
term
\equ(inhomog).

We next consider the set of functions satisfying
$$
\sup_{0\le t\le T}e^{\rho(T-t)}\sup_{x}|v(t,x)|\,\le\, \eta^{2}~,
$$
with the associated metric.
A standard argument using the bound
\equ(contra) shows that in Eq.\equ(vequ)
we have a contraction (for
$\eta$ small enough, independent of $t$, $T$) in this
space and therefore a unique solution $v$ for the
Eq.\equ(vequ). Furthermore, the asserted bounds of \clm(lower1) follow
at once. We leave the (trivial) details to the reader. The proof of
\clm(lower1) is complete.

We now come to the proof of
convergence of $\Theta^Tf_{-T}$, as $T\to\infty $.
We shall show that the derivative of this quantity is integrable in $T$.
We recall that if we have a vector field
$X$ with  flow $\phi_{t}$ then
$$
{d\over dt}\phi_{t}(x)\,=\,\D\phi_{t}[x]X(x)~.
$$
We use throughout the notation $\D F[x]$ for the derivative of $F$
evaluated at $x$; this is usually an operator.
In our case, we get
$$
\eqalign{
{d\over
dT}\Theta^{T}(f_{-T})\,&=\,\D\Theta^{T}[f_{-T}]\cr
&\cdot\bigg((1+i\alpha)\partial_{x}^{2}
f_{-T}+f_{-T}-(1+i\beta)f_{-T}|f_{-T}|^{2}-(1+i\alpha)\partial_{x}^{2}f_{-T}
-f_{-T}\bigg)\cr
\,&=\,
-(1+i\beta)\D\Theta^{T}[f_{-T}]\big(f_{-T}|f_{-T}|^{2}\big)~.
}
\EQ(lower1)
$$
We want to prove that this quantity is integrable over $T$. For this
purpose we have to control the linear operator
$\D\Theta^{T}[f_{-T}]$.
\CLAIM Lemma(lower2) We have the inequality
$$
\|\D\Theta^{T}[f_{-T}]w\|_\infty \,\le\, \OO(1) e^{T(1+\OO (\eta))}\|
w\|_\infty ~.
\EQ(lower2a)
$$

\PROOF It
is easy to verify that  $\D\Theta^{T}[f_{-T}]w_0$
is given as the value at time $T$ of the solution of the
linear equation
$$
\partial_{t}w\,=\,(1+i\alpha)\partial_{x}^{2}w+w+R_\beta ~w+S_\beta
~\overline w~,
\EQ(DT)
$$
with initial condition $w(t=0,\cdot)=w_0(\cdot)$. The coefficients
$R_\beta $
and $S_\beta $ are given by
$$
R_\beta (t,x)\,=\,-2(1+i\beta)|\Theta^{t}(f_{-T})(x)|^2~,
$$
and
$$
S_\beta (t,x)\,=\,-(1+i\beta)\left(\Theta^{t}(f_{-T})(x)\right)^{2}~.
$$
The assertion of \clm(lower2) follows now, using a contraction
argument, as in the study of Eq.\equ(vequ),
from \clm(lower1) and the previous formula. The details are again left
to the reader.

As a consequence, combining the inequalities
\equ(firstlinfty) and \equ(lower2a),
the right  hand side of Eq.\equ(lower1) is exponentially
small in $T$ and therefore integrable and we have a limit.
So our map $S$ is well-defined by
$$
S(f)\,=\,\lim_{T\to\infty}\Theta^{T}(f_{-T})\,=\,
f+Z(f)~,
$$
where
$$
Z(f)\,=\,\lim_{T\to\infty}\int_{0}^{T}\kernint dt\,{d\over dt}\Theta^{t}(f_{-t})~,
$$
and in fact we have proven that this last term is of order $\eta^{2}$
(in reality $\eta^{3}$). This completes the proof of the first part of
\clm(superlower). It remains to prove that it is
Lipshitz and to estimate its Lipshitz constant in $\Linfty$. This
will be done in the next subsection, together with some even more
detailed information on $Z$ which we need later.

\SUBSECTION Proof of the second part of \clm(superlower)

In this subsection, we prove the second part of \clm(superlower), in
fact even more. We first need some notation:
\REMARK It will be more convenient to work with the
intervals $[-L,L]$ instead of $[-L/2,L/2]$ as in the earlier sections.
We shall use the following notations:
$$
\eqalign{
\B\,&=\,[-L,L]~,\cr
\S\,&=\,[-L+\ell,L-\ell]~,\cr
\S'\,&=\,[-L+\ell/2,L-\ell/2]~,\cr
\S''\,&=\,[-L+\ell/4,L-\ell/4]~,\cr
\BS\,&=\,[-L,-L+\ell)\cup (L-\ell,L]~.\cr
}
$$
These letters stand for ``big'' and ``small.''
Our result is
\CLAIM Proposition(lower3new) The function $Z$ is Lipshitz continuous in $f$
in a neighborhood of $0$ in $\EE_b(\eta)$, $b=1/3$,
with a Lipshitz constant ${\OO }(\eta)$:
$$
\|Z(f)-Z(f')\|_{\Linfty(\real)}\,\le\, \OO
(\eta)\|f-f'\|_{\Linfty(\real)}~.
\EQ(Lipshitz)
$$
Moreover, for $\ell\ge 1/\epsilon$ and $L$ large enough, one has the inequality
$$
\|Z(f)-Z(f')\|_{\Linfty(\S)}\,\le\, \OO (\eta)\|f-f'\|_{\Linfty(\B)}
+\OO (\epsilon^{2})\|f-f'\|_{\Linfty({\bf R})}~.
\EQ(lower2)
$$

Clearly, this result states more than what is asserted in
\clm(superlower), and thus, proving \clm(lower3new) will at the same
time
complete the proof of \clm(superlower).

\PROOF Using Eq.\equ(lower1), we have the expression
$$
Z(f)\,=\,\lim_{T\to\infty} Z_{T}(f)~,
$$
where
$$
Z_{T}(f)=-(1+i\beta)\int_{0}^{T}\kernint dt\,
\D\Theta^{t}[f_{-t}]\left(f_{-t}|f_{-t}|^{2}\right)~.
$$
To prove the first part of \clm(lower3new), we would like to obtain
a bound uniform in $T$ on the differential of $Z_T(f)$ with respect to
$f$. Due to the presence of the absolute value, this function is not
differentiable in $f$. One should therefore consider the expression
obtained by taking the real and imaginary parts (note that we are only
dealing with the values on the real axis and analyticity is
not used in the following argument). To make the exposition
simpler  we will only explain the proof for the real Ginzburg-Landau
equation (the field is real and $\alpha=\beta=0$), and for a space
dimension equal to one, but the general case only presents notational
complications.

We have then, since we assume $\beta =0$,
$$
Z_{T}(f)\,=\,-\int_{0}^{T}\kernint dt\,
\D\Theta^{t}[f_{-t}]\left(f_{-t}^{3}\right)~.
$$
From this formula we have
$$
\eqalign{
\D Z_{T}[f] w\,&=\,-\int_{0}^{T}\kernint dt\,
\D^{2}\Theta^{t}[f_{-t}](f_{-t}^{3},w)
\cr
\,&-\,3\int_{0}^{T}\kernint dt\,
\D\Theta^{t}[f_{-t}]\left(f_{-t}^{2}(\D f_{-t})w\right)\,\equiv\, X_1+X_2~.\cr
}
\EQ(deriv)
$$
The second term $X_2$ is easier to handle
and we first prove both Eq.\equ(Lipshitz) and \equ(lower2) for the
contributions coming from this term.
Since $f_{-t}$ is linear in $f$
we have
$$
(\D f_{-t})w\,=\,\Theta_{0}^{-t}w\,=\,w_{-t}~.
$$
Using \clm(lower2), and Eq.\equ(firstlinfty), the integrand is bounded
by
$$
\|\D\Theta^{t}[f_{-t}]\left(f_{-t}^{2}w_{-t}\right)\|_\infty\,\le\,
\OO(1)e^{t(1+\OO(\eta))}
\OO(\eta)e^{-2(1-c^2)t}\OO(1)e^{-(1-c^2)t}\|w\|_\infty~,
\EQ(exponential)
$$
and therefore we get a bound for the integral which is of the form
$$
\|3\int_{0}^{T}\kernint dt\,
\D\Theta^{t}[f_{-t}]\left(f_{-t}^{2}w_{-t}\right)\|_\infty \,\le\,
\OO(\eta)\|w\|_\infty~,
$$
which shows that the contribution from $X_2$ to Eq.\equ(Lipshitz) is
of the desired form, by linearity.

We now come to the
localized bound Eq.\equ(lower2) for the contribution coming
from the term $X_2$. It is enough to assume $T$ large enough
and for example $T>t_0\log(1/\epsilon)$. Using the
exponential estimates of Eq.\equ(exponential),
we have
for a large enough constant $t_0$
(independent of $\epsilon$ small enough),
$$
\|3\int_{t_0\log\epsilon^{-1}}^{T}\kernint dt\,
\D\Theta^{t}[f_{-t}]\left(f_{-t}^{2}w_{-t}\right)\|_{\infty}
\,\le\,\OO(\epsilon ^2)\|w\|_{\infty}~.
$$
For the other part of the integral, from $0$ to $t_0\log\epsilon^{-1}$,
we proceed as in the proof of \clm(lower2).
We want to bound
$$
X_{2,+}\,=\,\int_{0}^{t_0\log\epsilon^{-1}}\kernint dt\,
\D\Theta^{t}[f_{-t}]\left(f_{-t}^{2}w_{-t}\right)~.
$$
In particular we will control the solution of the equation
$$
\partial_t{v}\,=\,\partial_{x}^{2}v+v+Rv~,
$$
where $R=\OO(\eta^{2})$.  Note that this is very similar to the
estimate in \clm(3), but the proof is more delicate.

We can write an integral equation, namely
if $K_{t}$ is the heat kernel (associated with the Laplacian),
we have
$$
v_{t}\,=\,e^{t}K_{t}\star v_{0}+\int_{0}^{t}\kernint ds\,e^{t-s}K_{t-s}\star
(R_{s}v_{s})~.
\EQ(vvequ)
$$
It is now convenient to define, as in the proof of \clm(3),
$$
y_{t}\,=\,e^{-t(1+\eta)}v_{t}~,
\EQ(ydef)
$$
and to prove uniform bounds in $t$ for $y_{t}$. This leads to the
integral
equation
$$
y_{t}=K_{t}\star v_{0}+\int_{0}^{t}\kernint ds\,e^{-\eta(t-s)}K_{t-s}\star
(R_{s}y_{s})~.
\EQ(yequ)
$$
In particular, if we consider this equation in the space of functions
bounded in space and time, the last term gives an operator of norm
$\OO(\eta)$ because $R=\OO(\eta^{2})$. Therefore we can solve
this
equation for $\eta$ small by iteration ({\it i.e.}, the Neumann series
converges). This is really the proof of \clm(3).
We are going to use this idea in a slightly more subtle way,
taking advantage of the decay properties of the heat kernel. We first
choose a number $c_{1}>0$ large enough, basically $c_{1}^2/t_0\gg1$, where
$t_0$ was defined above.
We then choose an integer $n$ such that
$$
{(\log\epsilon)^{2}\over \log\eta^{-1}}\,\ll\,n~,\quad{\rm and}\quad
nc_1\log(1/\epsilon ) \,\le\, \ell/4~.
$$
Clearly, for our choice of $\ell\ge1/\epsilon$, and
since $\eta$ is a fixed (but small) constant, we can choose $n$ for
example of order $\bigl (\log(1/\epsilon )\bigr )^3$, if $\epsilon>0 $
is small enough.

We next define a sequence of domains for $0\le j\le n$ by
$$
\S'_{j}\,=\,
[-L+\ell/2-jc_{1}\log(1/\epsilon),L-\ell/2+jc_{1}\log(1/\epsilon)]~.
$$
Note that the distance between $\S'_{j}$ and the complement of $[-L,L]$ is
at least $\ell/4$ (for $\epsilon$ small enough), that
$\S'_j\subset\S'_{j+1}$,
that $\S_0=\S'$ and that $\S_n\subset[-L+\ell/4,L-\ell/4]=\S''$.
Using the integral equation Eq\equ(yequ)
and $t\le t_0\log\epsilon^{-1}$ we find, upon
splitting the convolution integrals in the space variable, and
writing
$t^*= t_0 \log \epsilon ^{-1}$:
$$
\eqalign{
\sup_{t\in[0,t^*]}\|y_{t}\|_{\Linfty(\S'_{j})}
\,\le\, &\|v_{0}\|_{\Linfty({\bf B})}+{\OO }(\epsilon^{2})
\|v_{0}\|_{\Linfty({\bf R})}\cr
+{\OO (\eta)}\sup_{t\in[0,t^*]}
\|y_{t}\|_{\Linfty(\S'_{j+1})}&+{\OO }(\epsilon^{2})
\sup_{t\in[0,t^*]}
\|y_{t}\|_{\Linfty({\bf R})}~.
}
\EQ(vbound)
$$
For example, the term $K_t\star v_0$ is bounded as follows: Writing
$t^*= t_0 \log \epsilon ^{-1}$we have
$$
\eqalign{
\sup _{t\in[0, t^*]}\sup_{x\in \S'_j} |\bigl (K_t\star
v_0\bigr )(x)|
\,&\le\,
\sup _{t\in[0,t^*]}\sup_{x\in \S'_j}\int_{z\in
\B\cup {(\real\setminus \B)}} \kern -1em dz\, |K_t(x-z) v_0(z)|\,\equiv X_\B +
X_{\real\setminus \B}~.\cr
}
$$
The term $X_\B$ leads to the bound $\|v_0\|_{\Linfty(\B)}$, since the
integral of $|K_t|= K_t$ equals 1. Using $K_t(z)\le
2^{1/2}e^{(z^2/(2t))}K_{2t}(z)$, the term $X_{\real\setminus\B}$ is
bounded by the supremum of $v_0$ times
$$
\eqalign{
\sup_{t\in[0, t^*]}&\int _{|z|>\ell/2- j c_1 \log(1/\epsilon )}\kern -1em dz\,
K_t(z)\cr\,\le\,
\sup_{t\in[0, t^*]}\,\,\,\sup_{|x|>\ell/2- j c_1
\log(1/\epsilon)}& \OO(1) \exp\bigl (-\const x^2/(2t)\bigr)\cdot \int_\real dz\,(K_{2t}(z))~.\cr
}
$$
Our choice of $n$ and $c_1$ implies that $x^2/t\ge \log(1/\epsilon^2)$
(in fact a much better bound holds here, but later, when we iterate
the argument, we shall use a
bound which essentially saturates this inequality)
and thus the bound of the first term in Eq.\equ(yequ) follows. The bound
on the second term follows using the same techniques and the
contraction mapping principle as in our treatment of Eq.\equ(vequ),
and using that $R_s=\OO(\eta^2)$ to compensate for a factor of
$\eta^{-1}$ which comes from the bound on the $s$-integral.

Using the estimate on the whole line (\clm(3)),
we conclude that the last term in \equ(vbound) is of the same size
as the second term and we get
$$
\sup_{t\in[0,t^*]}\|y_{t}\|_{\Linfty(\S'_{j})}
\le  \|v_{0}\|_{\Linfty({\bf B})}
+{\OO }(\epsilon^{2})
\|v_{0}\|_{\Linfty({\bf R})}+
{\OO (\eta)}\sup_{t\in[0,t^*]}
\|y_{t}\|_{\Linfty(\S'_{j+1})}~.
$$
We now iterate $n$ times this inequality (and here we only get a bound
$x^2/t>\log(1/\epsilon ^2)$ which comes from the lower bound on the
separation of $\real\setminus \S'_{j+1}$ from $\S'_j$) to obtain an estimate on
$\S_{0}={\S'}$. Since we have chosen the constant
$n$
such that
$\eta^{n}={o}(\epsilon^{2})$, we find
$$
\sup_{t\in[0,t^*]}\|y_{t}\|_{\Linfty({\bf S'})}
\le  {\OO }(1)\|v_{0}\|_{\Linfty({\bf B})}
+{\OO }(\epsilon^{2})
\|v_{0}\|_{\Linfty({\bf R})}~.
$$
We can now undo the effect of the exponential of Eq.\equ(ydef). If we
furthermore replace $v_0$ by the initial data
$f_{-t}^{2}w_{-t}$, and use the information we have on
$f_{-t}^{2}w_{-t}$,
we get the bound for this part of the integral:
$$
\|X_{2,+}\|_{\Linfty(\S')}\,\le\,\OO(\eta)\|w\|_{\Linfty(\B)}
+\OO(\epsilon ^2)\|w\|_{\Linfty(\real)}~.
$$
Since $\S'\supset\S$, this is the desired bound, and we have completed
the bound on $X_2$.

We finally consider the term $X_1$ of Eq.\equ(deriv).
Here, we estimate
$$
\D^{2}\Theta^{t}[f](w_{1},w_{2})~.
$$
Again, this is a function $z$ which is a solution of
$$
\partial_{t}z=\partial_{x}^{2}z+z-3\bigl (\Theta^{t}f\bigr
)^{2}z-6\Theta^{t}f\cdot 
\D\Theta^{t}[f]w_{1}\cdot\D\Theta^{t}[f]w_{2}~,
\EQ(DT2)
$$
with initial data $z=0$,
which is the analog of the Eq.\equ(DT) which we found for the first derivative.
Its estimate is analogous to the previous one. To deal with
the localization problem for the non-homogeneous term in \equ(DT2), we
now exploit
that the bound on $X_2$ was done on a region $\S'$ which is larger
(by $\ell/2$) than the region $\S$ on which we really need the bounds.

Details are left to the reader.

\LIKEREMARK{Interpretation}The inequality Eq.\equ(lower2) serves to
localize the bounds of the previous subsection.
If $\epsilon$ is small enough (depending only on the bounds \clm(superlower)
on the derivative of $Z$ which are
global), we have for any two functions $f$, $f'$ in $\EE_b(\eta)$ the
inequality
$$
\|Z(f)-Z(f')\|_{\infty}\,\le\, \OO (\eta)\|f-f'\|_{\infty}~.
$$
Therefore,
$$
\|S(f)-S(f')\|_{\infty}\,\ge\, (1-\OO (\eta))\|f-f'\|_{\infty}~.
$$
Basically, we want to
use Eq.\equ(lower2) to show that if $f$ and $f'$ differ by at least
$\epsilon$ somewhere on
$\B$ this implies that $S(f)$ and $S(f')$ differ by at least
$\epsilon/2$ somewhere on $\S$.
While this is not true in general, we will see in the next section
that it must be true for enough functions among those which form the
centers of the balls which cover $\GG$.
This will be exploited in the next subsection.

\SUBSECTION Proof of \clm(lowerbound)

The idea of the proof is to show
that because $S(f)\approx f$ and because the \ee/ of
the set $\EE_b(\eta_*)$
of $f$ is $\OO\bigl (\logtwo(1/\epsilon )\bigr )$, the same will
hold for the set $S\bigl (\EE_b(\eta_*)\bigr )$. Here, and in the
sequel we fix $\eta_*$ to the value found as a bound in \clm(superlower).
Basically, we are going to show that if $\|f-f'\|_\infty \ge \epsilon
$, then not only
$$
\| S(f) - S(f') \|_\infty  \,\ge\,\epsilon /2~,
\EQ(easy)
$$
but also that we can find enough functions for which
$\sup_{x\in\B}| f(x)-f'(x)|>\epsilon $ and
$$
\sup_{x\in\S} | S(f)(x)-S(f')(x) | \,\ge\, \epsilon /4~.
\EQ(harder)
$$
Here, we shall choose $\ell\ge 1/\epsilon $.

Note that we cannot prove \equ(harder) for {\em individual} pairs of
functions, but only for a (large enough) subset of them. The mechanism
responsible for that is a ``crowding lemma'' in the
following setting: Let $\SS$ be a set of $N\gg 1$ functions which are pairwise
at a distance at least $\alpha$ from each other, when considered on
a set $I_{\rm big}$ which is a
finite union of intervals. Let $I_{\rm small}$ be another finite union of
intervals 
contained in $I_{\rm big}$. 
\CLAIM Lemma(pairs) Under the above assumptions at least one of the
following alternatives holds:
\item{--}At least $N^{1/2}/2$ functions in $\SS$ differ pairwise
by $\alpha$ on $I_{\rm big}\setminus I_{\rm small}$.
\item{--}At least $N^{1/2}$ functions in $\SS$ differ pairwise
by $\alpha/3$ on $I_{\rm small}$.

\REMARK We can symmetrize the statement.
We formulate this as a corollary for further use:
\CLAIM Corollary(pairs2) Under the above assumptions at least one of the
following alternatives holds:
\item{--}At least $N^{1/2}/2$ functions in $\SS$ differ pairwise
by $\alpha/3$ on  $I_{\rm big}\setminus I_{\rm small}$.
\item{--}At least $N^{1/2}/2$ functions in $\SS$ differ pairwise
by $\alpha/3$ on $I_{\rm small}$.

\PROOF We first need the following auxiliary
\CLAIM Lemma(pairwise) Let $\EE$ be a set of $M^2>4$ points in a metric
space.
Assume
that for a given $\rho>0 $ we can find in $\EE$ no more than $M$ points
which are pairwise at a distance at least $\rho $. Then there is a
point $x_*$ in $\EE$ such that at least $M/2$ points of $\EE$ are within
a distance $\rho $ of $x_*$.

\PROOF Let $\EE_0$ be a maximal set of points in $\EE$ with pairwise
distance at least $\rho$. By assumption, the cardinality
of $\EE_0$ satisfies $|\EE_0|\le M$. Adding any point
$x_0\in\EE\setminus\EE_0$ to $\EE_0$, we
can find a point $x'_0\in\EE_0$ such that $d(x_0,x'_0)<\rho $, where $d$
is the distance. We continue in this fashion with every point $x_j$ of
$\EE\setminus\EE_0$, finding a partner $x'_j$ in $\EE_0$ with
$d(x_j,x'_j)<\rho $. There are thus $|\EE\setminus\EE_0|=M^2-M$
choices of $x'_j$. But
since there are at most $M$ points in $\EE_0$, there must be at least
one point in $\EE_0$ which has at least $(M^2-M)/M$ partners.
Clearly, this point can be chosen as $x_*$. Since $(M^2-M)/M>M/2$,
the proof of
\clm(pairwise) is complete.
\LIKEREMARK{Proof of \clm(pairs)}We assume that the second alternative
does not hold and show that then the first must hold. If the second
alternative does not hold, then we can apply \clm(pairwise) with
$(M+2)^2> N\ge (M+1)^2$ and $\rho=\alpha/3 $
on the set $\SS$ of functions with the sup norm on $I_{\rm small}$ and
conclude that there is a function, $f^*$, such that on $I_{\rm small}$ we
can find $M/2$ others within distance at most $\alpha/3$ from
$f^*$. Call those functions
$f_{i}$ ($i=1,\dots, K $ with  $K \ge M$).
Therefore,
$$
\sup_{x\in I_{\rm small}}|f_{j}(x)-f_{j'}(x)|\,<\,2\alpha/3~,
$$
for all pairs $j,j'\in\{1,\dots,K \}$.
This implies that these $M/2$ functions $f_{i}$ must differ
pairwise by
at least $\alpha$ on $I_{\rm big}\setminus I_{\rm small}$ since
they have to differ pairwise by $\alpha$ on the whole interval
$I_{\rm big}$. 
The proof of \clm(pairs) is complete.

With these tools in place, we can now start the proof of
\clm(lowerbound) proper.
We first make precise the limiting process in the definition of
$H_\epsilon\bigl (\EE_b(\eta_*)\bigr )$.
Using the definition of $H_\epsilon$ we have the following
information
about the set $\EE_b(\eta_*)$: Let $N_{[-L,L]}(\epsilon )$ denote again
the minimum number of balls of radius $\epsilon $ in $\Linfty([-L,L])$
needed to cover $\EE_b(\eta_*)$ (restricted to $[-L,L]$). Then we know
that
$$
\lim_{\epsilon \to 0} {1\over \logtwo(1/\epsilon )}\lim_{L\to \infty}{
\logtwo N_{[-L,L]}(\epsilon )\over 2L} \,=\, {2b\over \pi}~.
$$
This leads to upper and lower bounds of the following form:
{\sl For every $\delta >0$ there is an $\epsilon (\delta )>0$
and for every
$\epsilon $ satisfying $0<\epsilon <\epsilon (\delta )$ there is an
$L(\delta ,\epsilon )$ such that for all $L>L(\delta ,\epsilon )$ one
has
$$
 \left ({1\over \epsilon }\right
)^{\sigma_*L(1-\delta )}\,\le\,N_{[-L,L]}(\epsilon )\,\le\, \left ({1\over \epsilon }\right
)^{\sigma_*L(1+\delta )}~,
\EQ(basic)
$$
where $\sigma_*= 4b/\pi$.}
Given $\delta >0$,
we pick $\epsilon $ and $L$ as above
and can find
therefore in $\EE_b(\eta_*)$ a set $\SS_1$ of
$$
N_1(\epsilon ,L)\,\ge\, \left ({1\over \epsilon }\right
)^{\sigma_*L(1-\delta )}~,
\EQ(sigma)
$$
functions which are pairwise at distance at least $\epsilon$ in
$\Linfty(\B)$.

\CLAIM Lemma(ss1) When $\ell\gg1/\epsilon  $ and $L\gg \ell$ on can
find
in $\SS_1$ a set $\SS_2$ of at
least $N_2={1\over 2}N_1^{1/2}$ functions which differ
pairwise by $\epsilon /3$ on $\Linfty(\S)$.

\PROOF We apply \clm(pairs2) with $I_{\rm big}=[-L,L]$ and $I_{\rm
small}=[-L+\ell,L-\ell]$ and with $N=[(1/\epsilon)^{\sigma_* L(1-\delta
)/2}]$ and $\alpha =\epsilon $.
If the conclusion of \clm(ss1) does not hold, then by \clm(pairs) we
can find $N_2$ functions which are pairwise at a distance at least
$\epsilon $ on $[-L,-L+\ell]\cup[L-\ell,L]$. Applying 
\clm(pairs2) with $I_{\rm big}=[-L,-L+\ell]\cup[L-\ell,L]$ and $I_{\rm
small}=[-L,-L+\ell]$ we conclude that in at least one of the intervals
$[-L,-L+\ell]$ and $[L-\ell,L]$ we can find
at least $N_3=\HALF N_2^{1/2}$ functions
which are pairwise at a distance $\epsilon /3$ when considered on that
interval. Since we are considering a subset of
$\EE_b(\eta_*)$,
we see by Eq.\equ(basic),
there can be no
more
than $N_4\equiv(1/\epsilon)^{\sigma_*(1+\delta ) \ell}$
such functions. Since $\delta>0$ is arbitrarily small and we have seen
that there are at least
$N_3$ such functions, we find for
$L/5>\ell(1+\delta)/(1-\delta)+1$ the inequality $N_3>N_4$. This is a
contradiction and the
proof of \clm(ss1) is complete.

Continuing the proof of \clm(lowerbound),
we take the set $\SS_2$ of $N_2$ functions among the initial
ones which differ pairwise at least by $\epsilon/3$ on $\S$. Note that
this is different from looking at functions which differ by $\epsilon/3$
only on that interval because in $\SS_2$ we have some information outside, namely
that the functions differ by at least $\epsilon $ when considered on $\B$.
We consider the different $S(f)$ for these functions. Assume first
that at least
$$
N_5\,\equiv\, {1\over 2}N_2^{1/2}=\OO\bigl (\epsilon
^{-L\sigma_*(1-\delta)/4}\bigr )
$$ of these $S(f)$ differ pairwise by at
least
$\epsilon/12$ on $\S$. This means that $N_5$ balls of radius
$\epsilon /25$ in $\Linfty(\S)$
do {\em not} cover the set $S(\SS_1)$.
In the terminology of [KT, pp 86--87], this means that the minimal
number of
points in an $\epsilon /25$-net is at least $N_5$. Thus the \ee/ per
unit length of
$S(\SS_1)$ is bounded below by $\OO\bigl (\logtwo(1/\epsilon )\bigr )$,
we have a lower bound and we are done, {\it i.e.}, \clm(lowerbound) is
proved in this case.

For the opposite case, we are going
to derive a contradiction, and this will complete the proof of
\clm(lowerbound) for all cases.
By \clm(pairwise),  with $\rho=\epsilon /12$, if we cannot find at least
$N_5$ of the $S(f_i)$ which differ pairwise by at least $\epsilon
/12$ on $\S$,
there
is an $f^{**}$ such that in a neighborhood of radius $\epsilon/36$
around
$S(f^{**})$  we can find at least $N_5$ of the
other $S(f)$. This implies that we have a sub-collection $\{f_{i}\}$
of $N_5$ functions for which
$$
\sup_{x\in\S}|S(f_{j})(x)-S(f_{j'})(x)|<\epsilon/36~,
$$
for all choices of $j$ and $j'$.
Therefore, by the definition of $S$ and $Z$ we have
$$
\sup_{x\in\S}|f_{j}(x)-f_{j'}(x)|<\epsilon/36+
\sup_{x\in\S}|Z(f_{j})(x)-Z(f_{j'})(x)|~.
$$
We now apply Eq.\equ(lower2)
to bound this quantity by
$$
\sup_{x\in\S}|f_{j}(x)-f_{j'}(x)|<\epsilon/36+{\cal
O}(\epsilon^{2})+\OO (\eta_*)\sup_{x\in\B}|f_{j}(x)-f_{j'}(x)|~.
$$
Now if
$$
\sup_{x\in\BS}|f_{j}(x)-f_{j'}(x)|\,\le\,
\sup_{x\in\S}|f_{j}(x)-f_{j'}(x)|~,
$$
the previous inequality implies
$$
\sup_{x\in\S}|f_{j}(x)-f_{j'}(x)|\,<\,(1+\OO (\eta_*))^{-1}(
\epsilon/6+\OO (\epsilon^{2}))~.
$$
Combining the last two inequalities
we have
$$
\sup_{x\in\B} |f_{j}(x)-f_{j'}(x)|\,<\,(1+\OO(\eta_*))^{-1}(
\epsilon/36+\OO (\epsilon^{2}))~,
$$
and we have a contradiction since the distance should be at least
$\epsilon$. (It is here that we use the additional information we have
on the set $\SS_2$ of $N_2$ functions constructed in \clm(ss1).)
Therefore we conclude that
$$
\sup_{x\in\BS}|f_{j}(x)-f_{j'}(x)|\,>\,
\sup_{x\in\S}|f_{j}(x)-f_{j'}(x)|~,
$$
but since the sup over the whole interval must be $\epsilon$ we conclude
that the sup on the l.h.s.~is at least $\epsilon$. Applying again
\clm(pairs2) we
can find among the 
$\{f_{i}\}$ at least ${1\over 2}N_5^{1/2}$ functions such that on
one of the intervals $[-L,-L+\ell]$ or $[L-\ell,L]$ of $\BS$
they are pairwise at a
distance at least $\epsilon/36$. As before this leads to a contradiction if
$L\gg \ell$ because there should be at most $\epsilon ^
{-\ell\sigma_{*}(1+\delta )}$ such functions.
The proof of \clm(lowerbound) is complete.

\LIKEREMARK{Acknowledgments}This work was supported in part by the
Fonds National Suisse.
\SECTIONNONR References
  
\eightpoint
\widestlabel{XXX}
\ref
\no C
  \by Collet, P. 
\paper Non linear parabolic evolutions in unbounded domains 
\inbook Dynamics, Bifurcations and Symmetries
\bybook P.Chossat, ed
\pages 97--104 
\publisher Nato ASI 437, New York, London, Plenum,
\yr 1994
\endref
\ref
\no CE
  \by Collet, P. and J.-P. Eckmann
  \paper The time-dependent amplitude equation for the Swift-Hohenberg problem
  \jour Commun. Math. Phys.
  \vol 132
  \pages 139--153
  \yr 1990
\endref
\ref
  \no ER
  \by Eckmann, J.-P. and D. Ruelle
  \paper Ergodic theory of chaos and strange attractors
  \jour Rev. Mod. Phys.
  \vol 57
  \pages 617--656
  \yr 1985
\endref
\ref
\no G
  \by Gallay, Th.
  \paper A center-stable manifold theorem for differential equations in Banach spaces
  \jour Commun. Math. Phys.
  \pages 249--268
  \vol 152
  \yr 1993 
\endref	
\ref 
  \no GH
  \by Ghidaglia, J.M. and B. H\'eron
  \paper Dimension of the attractors associated to the Ginzburg-Landau
  partial differential equation
 \jour Physica
 \pages 282--304
 \yr 1987
 \vol 28D
\endref
\ref 
  \no HS
  \by Hocking, L.M. and K. Stewartson
  \paper On the nonlinear response of a marginally unstable plane
  parallel flow to a two-dimensional disturbance
 \jour Proc. R. Soc. Lond.
 \vol A. 326
 \yr 1972
 \pages 289--313
\endref
\ref
  \no H
  \by H\"ormander, L.
  \book The Analysis of Linear Partial Differential Equations
  \publisher Berlin, Heidelberg, New York, Sprin\-ger
  \yr 1983--1985
\endref
\ref
 \no KT
 \by Kolmogorov, A.N. and V.M. Tikhomirov
 \paper $\epsilon$-entropy and $\epsilon $-capacity of sets in
 functional spaces\footnote{${}^1$}{The version in this collection is
 more complete than the original paper of Uspekhi Mat. Nauk, {\bf 14},
 3--86 (1959).}
 \inbook Selected Works of A.N Kolmogorov, Vol III
 \bybook Shirayayev, A.N., ed.
 \publisher Dordrecht, Kluver
 \yr 1993
\endref 
\ref
  \no M
  \by Mattila, P.
  \book Geometry of Sets and Measures in Euclidean Spaces
 \publisher Cambridge, Cambridge University Press
  \yr 1995
\endref
\ref
\no MS
  \by Mielke, A. and Schneider G.
  \paper Attractors for modulation equations on unbounded
 domains---existence and comparison
  \jour Nonlinearity
  \vol 8
  \pages 743--768
  \yr 1995
\endref
\ref
  \no R
  \by Ruelle, D.
  \book Statistical Mechanics
  \publisher New York, Benjamin
  \yr 1969
\endref
\ref 
\no S
\by Schwartz, L.
\book Th\'eorie des Distributions
\publisher Paris, Hermann
\yr 1950
\endref
\ref
\no T
  \by Temam, R.
  \book Infinite-dimensional dynamical systems in mechanics and physics
  \publisher New-York, Springer-Verlag
  \yr 1988
\endref 
\bye